\documentclass[aps,prb,twocolumn,superscriptaddress,10pt]{revtex4-2}

\usepackage{dcolumn}
\usepackage{silence}
\usepackage{dsfont} 
\usepackage{bm}
\usepackage{dsfont}
\usepackage{epsfig}
\usepackage{amssymb} 
\usepackage{amsfonts} 
\usepackage{mathtools} 
\usepackage{dcolumn} 
\usepackage{graphicx} 
\usepackage{color}  
\usepackage{bm}  
\usepackage{comment} 
\usepackage{units}
\usepackage{textcomp} 
\usepackage{units}
\usepackage{textcomp} 
\usepackage{mathrsfs} 
\usepackage{array} 
\usepackage{natbib}
\usepackage{upgreek}
\usepackage{tabularx}
\usepackage{latexsym}
\usepackage{amsmath}
\usepackage{braket}
\usepackage{comment}
\usepackage[breaklinks=true,pdfborder={0 0 0}]{hyperref}
\def \blio{$\beta-$Li$_{2}$IrO$_{3}$}

\def \nio{Na$_{2}$IrO$_{3}$}
\def \lio{Li$_{2}$IrO$_{3}$}
\def \lro{Li$_{2}$RuO$_{3}$}

\def \rulio{Li$_{2}$Ru$_{x}$Ir$_{1-x}$O$_{3}$}
\def \runio{Na$_{2}$Ru$_{x}$Ir$_{1-x}$O$_{3}$}
\def \rucrcl{Ru$_{1-x}$Cr$_x$Cl$_3$}

\def \musr{$\mu$SR}

\usepackage{bigstrut}
\usepackage{placeins}
\usepackage[dvipsnames]{xcolor}

\usepackage{ccaption}

\begin{document}
\title{Anisotropic Kitaev Spin Glass in \texorpdfstring{\rulio}{Li2RuxIr1-xO3}}
\author{Mayia A. Vranas}
\email[]{mvranas@ucsd.edu}
\affiliation{Department of Physics, University of California, San Diego, California 92093, USA}

\author{Alejandro Ruiz}
\affiliation{Department of Physics, University of California, San Diego, California 92093, USA}

\author{Vikram Nagarajan}
\affiliation{Institute of Science and Technology Austria, 3400 Klosterneuburg, Austria}

\author{Erik Lamb}
\affiliation{Department of Physics, University of California, San Diego, California 92093, USA}

\author{Gerald D. Morris}
\affiliation{Centre for Molecular and Materials Science, TRIUMF, Vancouver, British Columbia, Canada V6T 2A3}

\author{Zahir Islam}
\affiliation{Advanced Photon Source, Argonne National Laboratory, Argonne, Illinois 60439, USA}

\author{Christie Nelson}
\affiliation{National Synchrotron Light Source II, Brookhaven National Laboratory, Upton, New York 11973, USA}

\author{Nobumichi Tamura}
\affiliation{Advanced Light Source, Lawrence Berkeley National Laboratory, One Cyclotron Road, Berkeley, California, 94720, USA}

\author{Benjamin A. Frandsen}
\affiliation{Department of Physics and Astronomy, Brigham Young University, Provo, Utah 84602, USA}

\author{James G. Analytis}
\affiliation{Department of Physics, University of California, Berkeley, California 94720, USA}

\author{Alex Frano}
\affiliation{Department of Physics, University of California, San Diego, California 92093, USA}

\date{\today}

\begin{abstract}
Kitaev iridates have emerged as an important class of spin–orbit-entangled quantum materials in which bond-directional exchange interactions generate strong magnetic frustration and unconventional correlated states. Rather than realizing simple ordered magnets, members of the $\alpha,\beta,\gamma-$\lio\ family exhibit fragile and highly anisotropic magnetic behavior, including incommensurate counter-rotating order, strong field sensitivity, and pressure-driven electronic reconstruction. These phenomena place the iridates in close proximity to the Kitaev quantum spin liquid (QSL) regime, while simultaneously revealing the importance of competing interactions, lattice distortions, and spin–orbit-assisted hopping processes beyond the ideal Kitaev limit. Understanding how chemical substitution perturbs these competing interactions provides a route toward probing the underlying frustrated magnetic state. Here, we study single crystals of $\beta-$\rulio\ with dilute Ru substitution, $x\lesssim10\%$. Through a combination of magnetometry, resonant elastic X-ray scattering, $ac$-heat capacity, and muon spin relaxation/rotation, we show that weak magnetic disorder continuously suppresses the incommensurate antiferromagnetic ground state without stabilizing an alternative long-range ordered phase. Instead, the system evolves into a bulk static spin glass characterized by slow relaxation, aging behavior, and frozen local magnetic fields. Despite the loss of long-range magnetic order, the glassy state retains substantial directional anisotropy, evidenced by enhanced response along the $b$-axis and strongly direction-dependent thermoremanent relaxation. Structural and reciprocal-space probes further indicate that Ru substitution leaves the hyperhoneycomb lattice largely undistorted while disrupting magnetic coherence, suggesting that disorder freezes spins within a still-active bond-directional exchange environment. In this picture, dilute Ru substitution provides a controlled pathway into an anisotropic Kitaev spin glass regime that preserves essential fingerprints of the underlying Kitaev exchange network.
\end{abstract}

\maketitle

\section{Introduction}

\par In quantum materials, the interplay of strong electronic correlations and spin-orbit coupling can generate local magnetic moments whose interactions are highly anisotropic and frustrated. One of the most prominent examples is the quantum spin liquid (QSL), a state in which magnetic moments avoid conventional long-range order and remain quantum mechanically entangled down to low temperature \cite{balents_spin_2010}. Among the proposed QSL frameworks, the Kitaev honeycomb model has attracted particular attention because it is exactly solvable, supports exotic fractionalized Majorana excitations, and may be realized in spin–orbit-entangled materials with bond-directional exchange interactions \cite{kitaev_anyons_2006, jackeli_mott_2009,chaloupkaKitaevHeisenbergModelHoneycomb2010a, matsuda_kitaev_2025}. Candidate Kitaev systems include RuCl$_3$ \cite{plumb_ensuremathalphaensuremath-mathrmrucl_3_2014}, the honeycomb iridates, A$_2$IrO$_3$ (A=Li,Na) and and the hydrogen-intercalated compound H$_3$LiIr$_2$O$_6$ \cite{singh_antiferromagnetic_2010,singh_relevance_2012, kitagawa2018, slagle2018}.

\par
\begin{figure*}[t]
  \centering
  \includegraphics[width=6in]{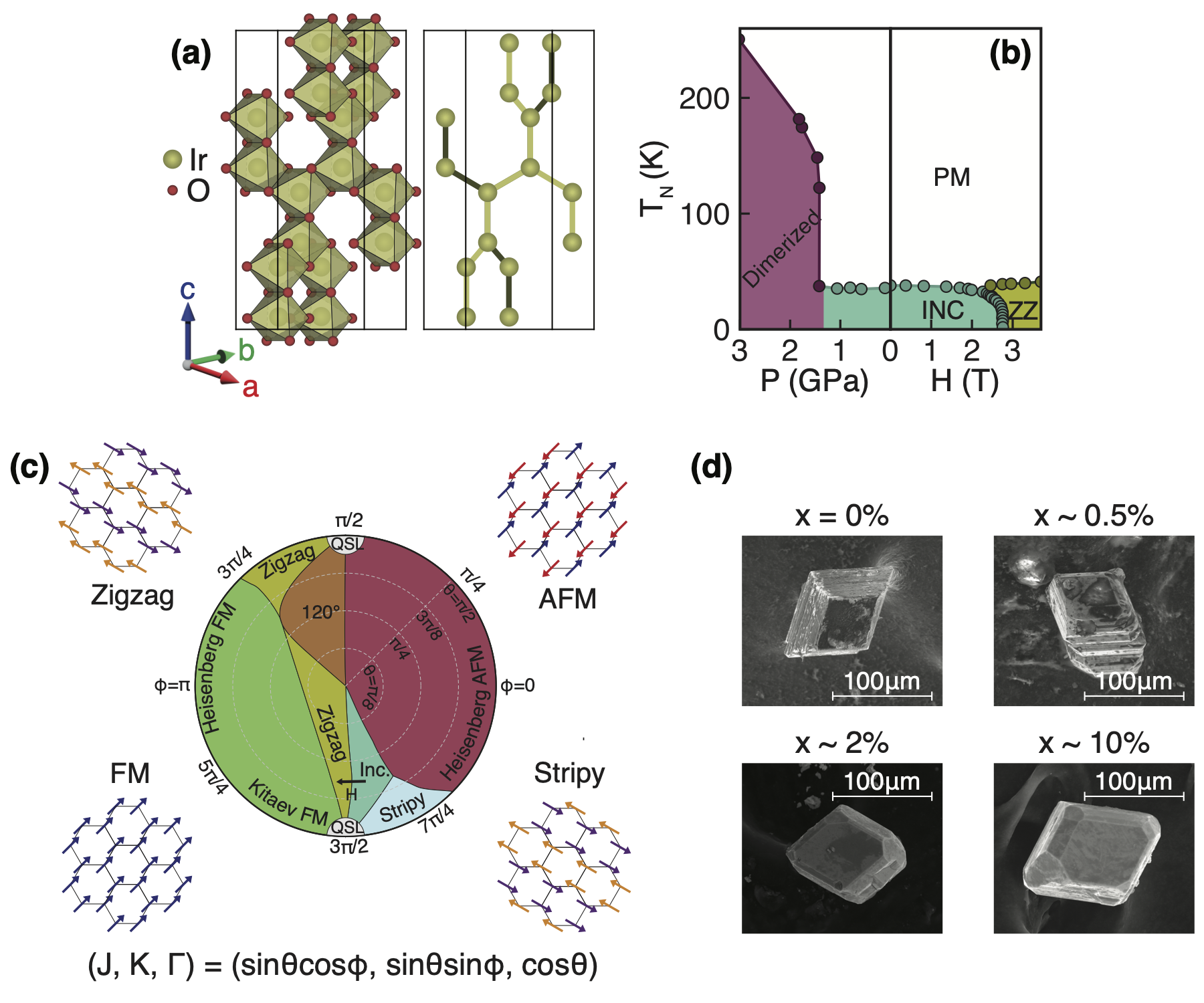}
  \caption{\textbf{Structure and Properties of \blio} \textbf{(a)} Crystal structure of \blio. Li ions are omitted for visibility. Ir-O octahedra form a twisted honeycomb structure (right panel). \textbf{(b)} Phase diagram for \blio\ as a function of pressure and applied field, from \cite{majumder_breakdown_2018,shen_interplay_2021,ruiz2017correlated}. \textbf{(c)} Phase diagram of the $JK\Gamma$ model, with schematics of different spin configuration. The black arrow shows the effect of applied magnetic field,  which moves the system from the incommensurate state to the zigzag state. Reproduced from \cite{chaloupkaKitaevHeisenbergModelHoneycomb2010a,rau_generic_2014}. \textbf{(d)} Scanning electron microscopy (SEM) images of four samples at various doping levels. Typical sample size ranges from $50-150\mu$m along the $b$-axis.}
  \label{fig:kitaev}
\end{figure*}

 In the honeycomb iridates, the essential ingredients for Kitaev physics arise from Ir$^{4+}$ ions confined within edge-sharing oxygen octahedra (or Ru$^{3+}$ ions within Cl$^-$ octahedra in the case of RuCl$_3$; Fig.~\ref{fig:kitaev}a). Crystal-field splitting combined with strong spin–orbit coupling produces spin–orbit-entangled $J_{eff}=1/2$ moments. The edge-sharing geometry of the octahedra generates bond-dependent exchange interactions along three inequivalent nearest-neighbor directions, commonly referred to as the Kitaev directions. In the ideal Kitaev honeycomb model, these bond-directional interactions yield an exactly solvable quantum spin liquid ground state whose excitations fractionalize into itinerant Majorana fermions coupled to emergent gauge fluxes \cite{kitaev_anyons_2006, jackeli_mott_2009,winterModelsMaterialsGeneralized2017a}.

\par Despite immense theoretical and experimental progress, most candidate Kitaev materials develop magnetic order, structural instabilities, or competing exchange interactions that drive them away from the ideal Kitaev limit \cite{biffin_noncoplanar_2014,biffin_unconventional_2014,williams_incommensurate_2016,liu_long-range_2011,sears_magnetic_2015}. In the iridates, strong spin–orbit-entangled hopping processes generate not only Kitaev interactions, but also Heisenberg ($J$), off-diagonal ($\Gamma$), and other longer-range exchange terms that stabilize complex magnetic states captured by the extended model \cite{jackeli_mott_2009,rau_generic_2014,rau2014a,katukuri2014}:
\begin{equation}
    H = \sum_{\braket{ij}\in\alpha\beta(\gamma)}[J\vec{S}_i\cdot\vec{S}_j + KS_i^\gamma S_j^\gamma + \Gamma(S_i^\alpha S_j^\beta+S_i^\beta S_j^\alpha)]
\end{equation}
Competition among these interactions stabilizes a wide variety of magnetic phases, including zigzag order, incommensurate spirals, field-induced correlated states, and quantum-disordered regimes \cite{rau_generic_2014,rau2014a,katukuri2014}.
\par As a result, the experimentally relevant problem has become how Kitaev-scale exchange interactions survive and reorganize under perturbations such as magnetic field, pressure, lattice distortions, and chemical substitution \cite{ruiz2017correlated,majumder_breakdown_2018, shen_interplay_2021,rolfs_spiral_2015,manni_effect_isoelectronic_2014,cao_evolution_2013,manni_effect_nonmagnetic_2014,mehlawat_fragile_2015,lei_structural_2014}. This issue has become especially important in 2D materials such as H$_3$LiIr$_2$O$_6$, where low-temperature excitation spectra may be consistent with Kitaev-like physics, but disorder-driven magnetic ground states and hydrogen-induced local inhomogeneity remain difficult to disentangle \cite{halloran2025,kitagawa2018,slagle2018}. More broadly, understanding how disorder modifies spin-orbit-entangled exchange networks has emerged as a key challenge across candidate Kitaev materials, since local structural disorder, random exchange, and bond inhomogeneity can strongly influence the stability and observable signatures of Kitaev-adjacent correlated states.

\par An important platform for exploring these effects is the hyperhoneycomb iridate \blio, shown in Fig.~\ref{fig:kitaev}a. In contrast to layered honeycomb systems, \blio\ realizes a three-dimensional network of edge-sharing IrO$_6$ octahedra that supports highly anisotropic and frustrated magnetic exchange interactions \cite{modicRealizationThreedimensionalSpin2014a}. At low temperatures, \blio\ develops an incommensurate counter-rotating spiral magnetic ground state proximate to the Kitaev regime \cite{biffin_unconventional_2014,halloran2022}. At higher temperatures, a distinct anomaly appears near $T_{\eta}$ under weak magnetic fields, which has been associated with the onset of strong bond-directional exchange correlations \cite{ruiz;prb20}. The coexistence of these two regimes makes \blio\ particularly sensitive to perturbations that modify the balance between Kitaev and non-Kitaev exchange interactions.

\par A wide range of studies have attempted to tune the magnetic state of \blio, producing the pressure–field–temperature phase diagram shown in Fig.~\ref{fig:kitaev}b. Relatively weak magnetic fields ($\mu_0H^\ast\sim2.4$T) suppress the incommensurate state and induce a correlated zigzag-like phase identified using resonant elastic X-ray scattering (REXS) \cite{ruiz2017correlated}. Pressure tuning studies, meanwhile, show that the transition temperature remains nearly unchanged up to $\sim1.4$GPa, above which Ir–Ir bond shortening drives dimerization and suppresses the low-temperature magnetic order \cite{majumder_breakdown_2018,shen_interplay_2021}. These results demonstrate that \blio\ is already proximate to multiple competing instabilities even before chemical substitution is introduced.

\par Chemical doping and disorder provide an additional route for perturbing these fragile exchange networks through modifications of local bonding, spin–orbit-assisted hopping, and magnetic disorder.. In $\alpha$-RuCl$_3$, for example, substitution on the Ru or halide sites destabilizes long-range magnetic order and can generate short-range correlated or glassy magnetic states \cite{do2018,kataoka2022,sato2024}. Similar behavior has been observed in Na$_2$IrO$_3$ and Li$_2$IrO$_3$, where nonmagnetic or magnetic substitution suppresses long-range order and stabilizes spin-glass-like behavior \cite{manni_effect_nonmagnetic_2014,mehlawat_fragile_2015,rolfs_spiral_2015}. Furthermore, disorder plays an important role in stabilizing this high-field phase of $\alpha$-RuCl$_3$, with the magnitude of anomalies in thermal hall conductivity significantly reduced in ultra-clean samples \cite{xing2025}. Theoretical studies have likewise predicted that impurities and disorder in Kitaev systems can produce a range of unconventional phenomena, including random-flux states, spin-glass phases, impurity-bound fluxes, and unconventional superconductivity in doped regimes \cite{willans_disorder_2010,andrade_magnetism_2014,cai_magnetic_2017,you_doping_2012,okamoto_global_2013,schmidt_topological_2018, chou2025}.

\par However, progress in understanding disorder's effects in \blio\ has been limited by the difficulty of synthesizing single crystals with dilute substitution levels. Previous studies of Ru-doped Li$_2$IrO$_3$ have focused primarily on powders or polycrystalline samples \cite{lei_structural_2014}, obscuring the directional anisotropies that are central to Kitaev exchange physics. In particular, the intermediate regime between long-range order and complete magnetic disorder remains largely unexplored, especially at low impurity concentrations where perturbations to the exchange network are subtle but potentially collective.

\par  In this work, we investigate the effect of Ru substitution in single crystal \rulio, in the dilute substitution regime ($x\lesssim10\%$), enabling a direct investigation of disorder-driven magnetism and anisotropic magnetic correlations at low impurity concentrations. Using SQUID magnetometry, $ac$-heat capacity, resonant elastic X-ray scattering (REXS), and muon spin relaxation/rotation (\musr), we demonstrate that dilute disorder does not simply suppress magnetic order, but rather stabilizes an anisotropic spin glass that preserves the bond-directional character of Kitaev exchange. These results reveal a previously inaccessible regime in which glassy freezing emerges from a highly frustrated, anisotropic magnetic manifold, rather than from conventional random exchange.

\begin{figure*}[t]
  \centering
  \includegraphics[width=\textwidth]{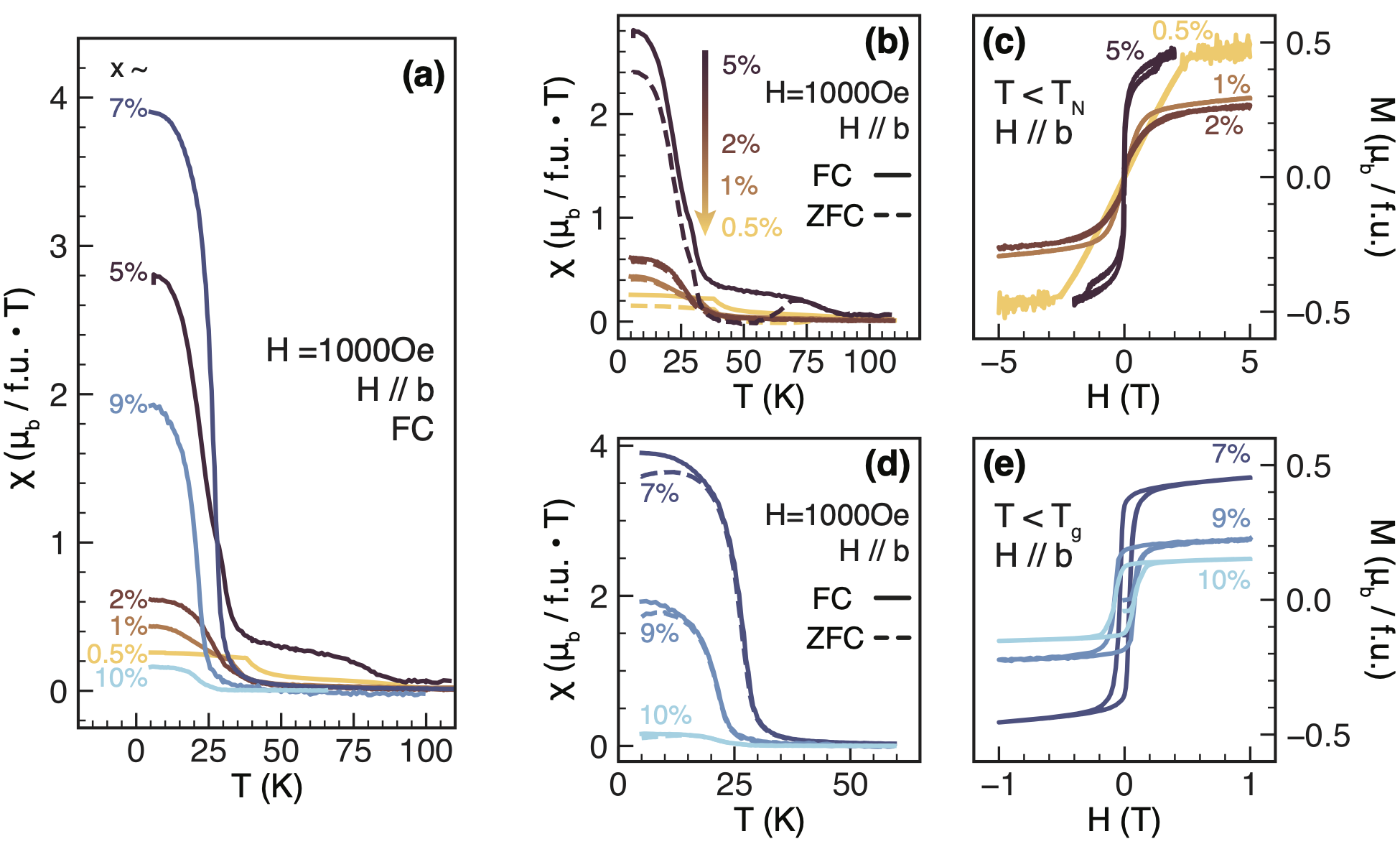} 
  \caption{\textbf{Magnetometry} \textbf{(a)} Field-cooled magnetic susceptibility for all samples, taken under a 1000Oe field applied parallel to the $b$-axis. \textbf{(b,d)} Field-cooled and zero-field cooled magnetic susceptibility in the low- and mid- doping regimes, respectively. A suppression of $T_N$ and $T_\eta$ and an increase in the total moment is seen, as expected with the doping of a $S=1$ ion. In the mid-doping regime, the AFM transition is replaced by a glassy transition. \textbf{(c,e)} Magnetization vs. applied field for low- and mid-doping regimes, respectively.}
  \label{fig:magnetometry}
\end{figure*}

\section{Methods}

\par High-quality single crystals of \rulio\ were grown by a two-step solid-state reaction. First, \lro\ was synthesized by reaction of Ru (99.9\% purity, BASF) and Li$_2$CO$_3$ (99.999\% purity, Alfa-Aesar) powders in the molar ratio of 1:1.05 for 84hr at 1010\textdegree C, followed by cooling  to room temperature at 7\textdegree C/hr. In the second step, Ir (99.9\% purity, BASF) and Li$_2$CO$_3$ (99.999 \% purity, Alfa-Aesar) powders were ground in a 1:1.05 stoichiometric ratio. \lro\ powder was added with mass calculated using:
\begin{equation*}
    m_{LRO}=m_{Li_2CO_3/Ir\,Powder}\cdot \frac{M_{LRO}}{M_{LIO}}\cdot\frac{x}{1-x}   
\end{equation*}
where $m$ is mass in grams, $M$ is molar mass, and $x$ is the desired Ru percentage. The ground powder was pressed into a pellet, reacted in an alumina crucible for 84hr at 1010\textdegree C, and cooled to room temperature at 7\textdegree C/hr. Single crystals of \rulio\ were then extracted from the reacted powder. Scanning electron microscopy was taken on a Quattro S ESEM, with crystal morphology reflecting sample quality (Fig. \ref{fig:kitaev}d). Sample volume varies on the order of $10^{-3}-10^{-2}$mm$^3$, with lengths of $50-100\mu$m along the crystalline $b$-axis.

\par Pristine \lro\ is believed to host an antiferromagnetic ordering of $S=1$ moments on the Ru$^{3+}$ lattice \cite{wangLatticetunedMagnetismRu2014}. However, other theoretical and experimental studies have shown the Ru$^{3+}$ sites have a tendency to dimerize, forming a valence bond liquid \cite{jackeliClassicalDimersDimerized2008,kimberValenceBondLiquid2014}. In this work on \rulio, we assume that dimerization is minimal, but further work is still needed.

\par Because direct compositional determination below $x\sim5\%$ Ru is not sufficiently reliable using methods such as energy-dispersive spectroscopy (EDS) on the available crystal sizes, the effective Ru concentrations discussed throughout this work are assigned phenomenologically using the nominal synthesis concentrations together with systematic trends in $T_N$, $T_\eta$, and magnetization behavior. All stated values are $x\lesssim10\%$ in accordance with results by Lei and colleagues \cite{lei_structural_2014}, whose lowest doping concentration of $x=10\%$ had a $T_c=3$K, lower than that of our highest dopant concentration. 

\par The precise local Ru valence state and resulting microscopic exchange modifications are not directly determined experimentally (see SI Sec. I). Structural characterization using Laue microdiffraction was conducted at the Advanced Light Source beamline 12.3.2 (Fig. 8). Structural peaks were indexed using X-ray Microdiffraction Analysis Software (XMAS) assuming the $Fddd$ (70) space group seen in pristine \blio. From this refinement, ratios of the lattice parameters $a/b,\,a/c,\,b/c$ were extracted and lattice parameters were estimated assuming constant unit cell volume. Single crystals were used for theromdynamic, magnetometry, and REXS measurements, whereas powders were used for $\mu$SR and powder X-ray diffraction (Fig. 9). All structural information derived is available in SI Sec. II.
\par Field- and temperature-dependent magnetization were studied using a Quantum Design MPMS3 SQUID. Both field-cooled (FC) and zero-field cooled (ZFC) magnetic susceptibility were studied, by cooling the sample with and without a magnetic field, and measuring through a heating cycle with an applied field of 1000Oe. $H$ was applied along the crystalline $b$-axis for all measurements, unless otherwise specified as in Fig. \ref{fig:anisotropy}.
\par The $ac$-specific heat $C_{ac}(T,H)$ measurements were conducted using a $\unit[16]{T}$ Cryogenic CFMS, which detects how the sample's temperature changes in response to an oscillating heater power. For this, a sample is placed over six thermocouples connected in series under a free-standing silicon nitride membrane $\sim$ 1$\upmu$m thick. An {\it ac} current with frequency $\omega$ is driven through an adjacent resistive heater, resulting in an oscillating power, $P_{ac}=\nicefrac{1}{2}I_o^2R\cdot(1+cos(2\omega t))$. The temperature of the sample V$_{ac}$ oscillates at a frequency $2\omega$ and can be used to calculate the $ac$-heat capacity:
\begin{equation}
C_{ac}\sim\frac{P_{ac}}{\omega \cdot V_{ac}}
\end{equation}
\par These measurements were performed in a low pressure He-4 gas environment ($\sim$10 mbar). The optimal frequency used in this experiment was $\unit[20]{Hz}$, necessary to ensure that the thermal link through the membrane and the gas can be ignored, and that the sample is heated homogeneously.
\par Magnetic REXS measurements were conducted at the Ir $L_3$ edge ($E=11.215$ keV) on single crystals of \rulio. These studies were conducted at both the Advanced Photon Source (APS) at beamline 6-ID-B and at the National Synchrotron Light Source II (NSLS II) at beamline 4-ID with vertical scattering geometry using a $\sigma$-polarized incident beam.
\begin{figure*}[t]
  \includegraphics[width=\textwidth]{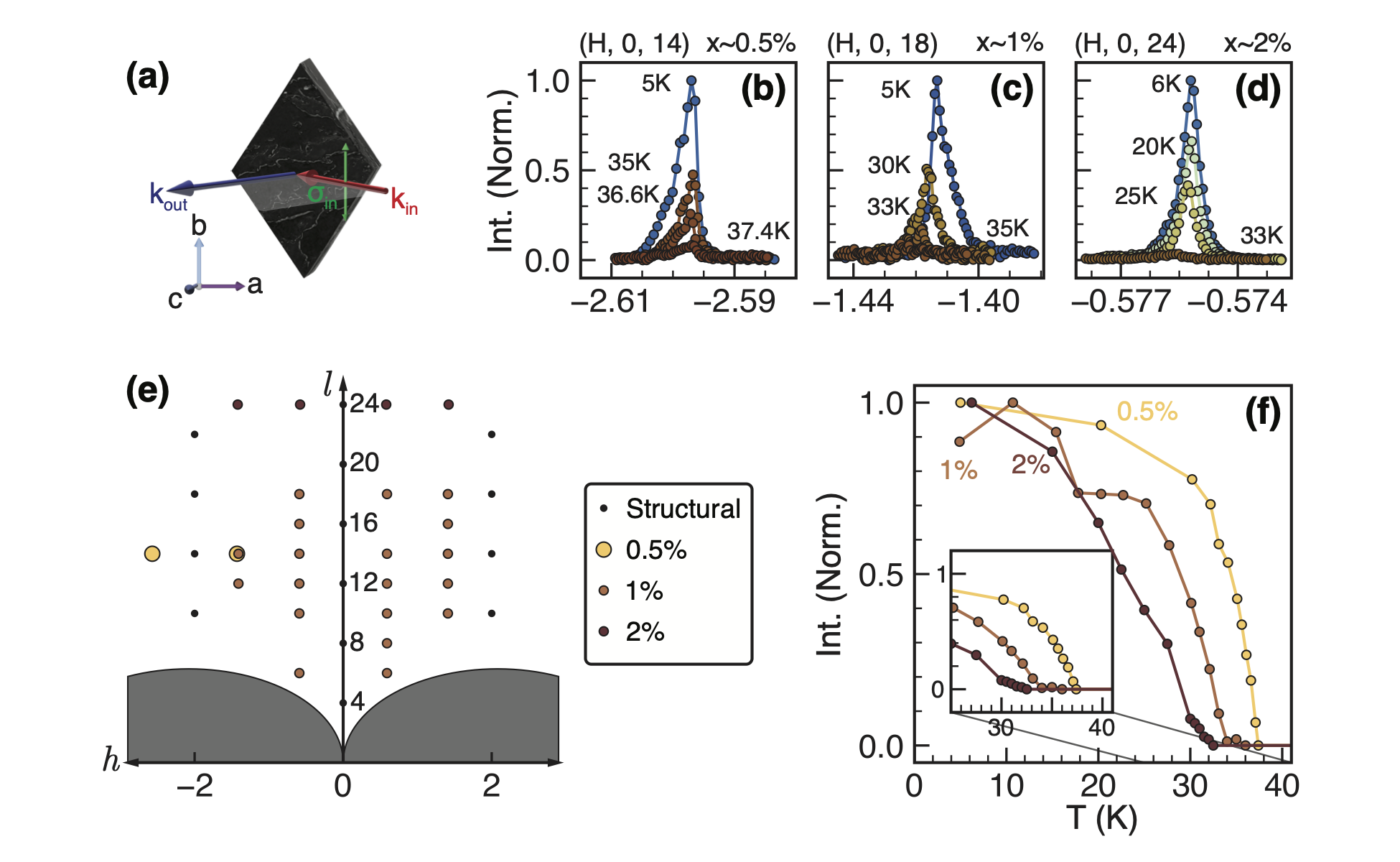}
  \caption{\textbf{Resonant Elastic X-ray Scattering (REXS)} \textbf{(a)} Diagram depicting the scattering geometry for REXS experiments. A vertical scattering geometry was used, such that the incident beam $k_{in}$ is polarized along $\sigma_{in}$ perpendicular to the scattering plane. \textbf{(b-d)} Peak profiles as a function of temperature for characteristic magnetic Bragg reflections seen in $x\sim0.5\%,\,1\%,\,2\%$ samples. Variations in alignment restricted which peaks were accessible, so different peaks were studied for each sample. Intensity is normalized to the peak intensity of the lowest measured temperature. \textbf{(e)} Mapping of the magnetic Bragg reflections $(h,0,l)$ seen in REXS experiments for different samples. Black circles indicate lattice reflections, while orange, yellow, and red circles correspond to signals seen for different dopants. \textbf{(f)} Self-normalized temperature dependence of the magnetic peak intensity. Inset shows systematic suppression of $T_N$ as a function of doping. $T_N$ found in scattering are consistent with those derived from magnetometry.}
  \label{fig:REXS}
\end{figure*}
\par Muon spin relaxation/rotation (\musr) experiments were conducted at TRIUMF Laboratory to locally investigate the emergent glassy behavior seen in \rulio. In a \musr\ experiment, spin-polarized, positively charged muons are implanted one at a time in the sample, where they typically come to rest at an interstitial position, undergo Larmor precession in any local magnetic field present at the stopping site, and subsequently decay into positrons after a mean lifetime of $\sim$2.2 $\mu$s. Pairs of detectors are placed on opposite sides of the sample to detect the positrons, which are emitted preferentially in the direction of the muon spin at the moment of decay. The experimental quantity of interest is the time-dependent asymmetric distribution of positrons detected by each detector in the pair, computed as 
\begin{equation}
    a(t) = (N_1(t)-N_2(t))/(N_1(t)+N_2(t))
\end{equation}
where $N_1(t)$ and $N_2(t)$ are the numbers of positrons incident on each detector at time $t$ after muon implantation. The asymmetry is proportional to the spin polarization of the muon ensemble projected onto the axis defined by the detectors, providing direct access to the local magnetic field distribution in the sample~\cite{hilli;nrmp22}. Powder samples with nominal compositions $x\sim0, 0.01, 0.05, 0.1,$ and $0.15$ were investigated using the LAMPF spectrometer on the M20D endstation. A helium gas flow cryostat with a base temperature of $\sim$1.8~K was used to control the temperature.

\section{Results}

\par In substituting less than $10\%$ Ru on the Ir sites of \blio, a trend emerges in which both the low-temperature $T_N$, marked by a slope change in magnetic susceptibility, and the high-temperature $T_\eta$, marked by a splitting of the FC-ZFC magnetic susceptibilities, are suppressed to lower temperatures. Because precisely measuring dopant levels below $10\%$ on small samples is challenging using existing methods (such as EDS and X-ray diffraction), we instead use these trends in the suppression of $T_N$, nominal Ru content used in synthesis, and comparison to previous studies on higher Ru content \cite{lei_structural_2014}. All stated nominal Ru concentrations are relative estimated values, and are not absolute. A full discussion of this determination is presented in SI Sec. I. Based on these trends, we present representative behaviors in two regimes: low doping ($0\%\lesssim x\lesssim5\%$) and mid-doping ($5\%\lesssim x\lesssim10\%$).

\begin{figure*}[t]
  \centering
  \includegraphics[width=\textwidth]{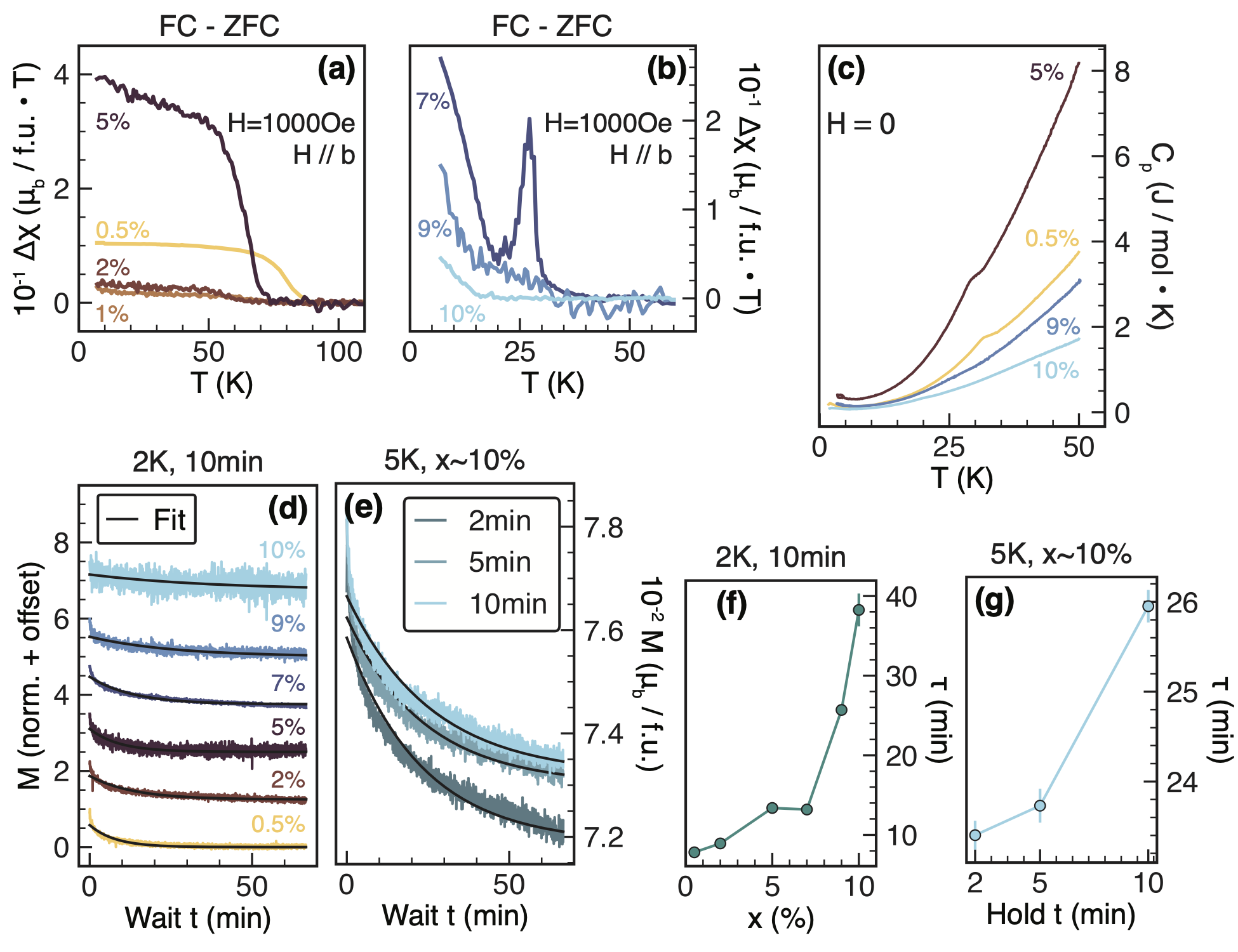}
  \caption{\textbf{Signatures of a Spin Glass in \rulio} \textbf{(a,b)} $\chi_{FC} - \chi_{ZFC}$ for low- and mid-doping regions, respectively. A shift from an order-parameter-like behavior, as seen in \cite{ruiz;prb20}, to a exponential growth is seen as the system moves into the glassy regime. \textbf{(c)} The $ac-$heat capacity for the low-doping regime shows a kink at $T_N$ characteristic of a second-order phase transition. This kink is suppressed, and only a broad feature is seen at $T_g$, characteristic of a spin glass. \textbf{(d)} Thermoremanent magnetization (TRM) as a function of doping after holding in a 1T field at 2K for 10min.  A longer relaxation time is seen for higher dopings. \textbf{(e)} TRM for $x\sim10\%$ at 5K with varying hold times at 1T. A longer relaxation time is seen for longer hold times. The dependence of the relaxation rate on the field-hold time is characteristic of aging behavior in spin glasses and reflects the slow evolution of the system through metastable magnetic configurations. \textbf{(f,g)} Characteristic timescale extracted from exponential fits of \textbf{(d,e)}. Higher doping levels correspond to longer characteristic timescales. In the glassy (mid-doping) regime, longer hold times also lead to longer characteristic timescales. Error bars are extracted from fit covariance.}
  \label{fig:spinglass}
\end{figure*}
\subsection{Bulk Magnetometry}

\par At low levels of Ru $x\lesssim5\%$, $T_{N,\eta}$ remain present, but are suppressed to lower temperatures (Fig. \ref{fig:magnetometry}(a,b)). $T_N$ is extracted via a kink in $\chi$ which is present for both FC and ZFC curves, while $T_{\eta}$ is apparent from the FC-ZFC splitting \cite{ruiz;prb20} (see SI Sec. I for further discussion). Additionally, the the kink field $H^*$, denoting a transition from INC to ZZ phases in pristine \blio\, is suppressed to lower values, and at $x\sim5\%$, a pinched hysteresis is present (Fig. \ref{fig:magnetometry}(c). Below $T_N$, for $1\%\lesssim x \lesssim 5\%$, rather than plateau, $\chi$ continues to increase, with its magnitude becoming greater for increased doping. This may indicate that the interactions responsible for stabilizing coherent long-range incommensurate order are being weakened, allowing a larger fraction of moments to align with the applied magnetic field. The pinched hysteresis observed in $x\sim5\%$ is consistent with a field-driven spin reorientation or spin-flop transition. The magnetometry data therefore support the interpretation that dilute Ru substitution destabilizes the coherent incommensurate antiferromagnetic state.
\par As $x$ is increased further to the mid-doping regime (Fig. \ref{fig:magnetometry}(d-e)), the behavior of the magnetic susceptibility and magnetization begin to change. The kink at $T_N$ is no longer present; instead, $\chi_{FC}$ increases until it plateaus at low temperatures, while $\chi_{ZFC}$ splits from $\chi_{FC}$ at the plateau and decreases. The FC-ZFC splitting temperature $T_g$ decreases with increasing doping. Hysteretic behavior is seen in the magnetization as a function of field which widens with increasing $x$; the saturation magnetization, meanwhile, begins to decrease with increasing $x$ in this regime.

\subsection{Resonant Elastic X-ray Scattering}

\par To explore the impacts in the low-doping regime on the magnetic propagation vector, REXS was conducted (Fig. \ref{fig:REXS}). These studies indicate that the antiferromagnetic propagation vector, which has a value of $\vec{q}=(0.567, 0, l)$ for even $l$ in pristine \blio, is still present with a small, non-systematic variation in the center position of $h$. A strong resonant enhancement is seen for each of these peaks, confirming their magnetic nature (Fig. 3-7). The magnetic peaks for each dopant level emerge at $T_N$, in agreement with values extracted from magnetic susceptibility (Fig. \ref{fig:REXS}b-d). The suppression of $T_N$ together with the small variation in the incommensurate propagation vector without movement of the structural Bragg reflections suggests that Ru substitution perturbs the balance of competing exchange interactions within the $JK\Gamma$ manifold. In particular, previous theoretical studies have shown that modifications of the off-diagonal $\Gamma$ interaction can alter the incommensurate ordering vector. A relative change in the off-diagonal exchange was indeed predicted to alter the incommensurate $\vec{q}$ \cite{rau_generic_2014}. No magnetic Bragg peaks associated with stripy or zigzag long-range order were observed within the explored region of reciprocal space, indicating that suppression of the incommensurate phase does not lead to stabilization of an alternative ordered magnetic state. While REXS studies were conducted on a single crystal of $x\sim10\%$, no magnetic peaks were apparent at the incommensurate, stripy, or zigzag sites. Taken together, the REXS results indicate that dilute Ru substitution progressively destabilizes the existing incommensurate magnetic state rather than reconstructing the system into a different long-range ordered phase.

\subsection{Signatures of a Spin Glass}
\par The combination of FC–ZFC irreversibility and the absence of coherent magnetic Bragg peaks in the mid-doping regime strongly suggests the emergence of a spin-glass ground state following the suppression of long-range incommensurate order. This can be seen more clearly by comparing $\Delta\chi=\chi_{FC}-\chi_{ZFC}$ for each doping. In the low-doping regime, order-parameter-like behavior is seen in $\Delta\chi$, in agreement with the behavior seen in pristine \blio in \cite{ruiz;prb20} which indicates the onset of Kitaev correlations (Fig. \ref{fig:spinglass}a). However, in the mid-doping regime, an exponential profile is seen, indicating that the origin of the FC-ZFC splitting is different from that in pristine \blio\ (Fig. \ref{fig:spinglass}b). A peak at $x\sim7\%$ is seen in $\Delta\chi$; The residual peak-like feature near ($x\sim7\%$) may indicate that short-range correlated behavior associated with the ($T_N < T < T_\eta$) regime of pristine \blio\ partially survives prior to the onset of fully developed glassy freezing.
\par Further evidence for a spin glass comes from the $ac-$heat capacity (Fig. \ref{fig:spinglass}(c)). In the low-doping regime, a discontinuity is seen in $C_p$ at $T_N$, as is expected for a second-order phase transition. However, in the mid-doping regime, the sharp thermodynamic anomaly at $T_N$ is replaced by a broad feature near $T_g$, consistent with glassy freezing rather than conventional long-range symmetry breaking.
\par Due to the small sample size, standard $ac$-magnetic susceptibility measurements were not possible to verify the glassy phase. Instead, time-dependent measurements were conducted to study the relaxation of the magnetic susceptibility, giving insight into the non-equilibrium dynamics of the system. Unlike ordered magnetic states, which typically relax rapidly after removal of an applied field, glassy magnetic systems exhibit slow, non-equilibrium relaxation due to the presence of metastable frozen configurations separated by broadly distributed energy barriers. This ``rugged'' energy landscape requires more traversal of phase space to find a low-energy configuration, so spin glasses are quantified by a thermoremanent magnetization (TRM) that persists to long timescales \cite{nordblad_time_1986}. To measure this TRM, the system was cooled without field below the glassy or antiferromagnetic transition, after which a field of 1T was applied for a fixed amount of time. This field was then removed, and the magnetization was measured for 75 min. Fig. \ref{fig:spinglass}d shows a scaled version of these curves, which have been normalized and offset to reflect how long it takes for the system to go from its initial to the lowest measured magnetization after 10min in a 1T field. The decay was fit to an exponential, from which the characteristic timescale $\tau$ was extracted:
\begin{equation}
    M = Ae^{-t/\tau}.
\end{equation}
Error bars were extracted from fit covariance. For samples in the  low-doping regime, a faster relaxation is seen, with $\tau\lesssim10$min, requiring less than ten minutes for the spins to relax into their ordered phase. As doping increases above $x\sim5\%$, a sharp upturn is seen in the relaxation time, with $\tau$ reaching 38 min for $x\sim10\%$ (Fig. \ref{fig:spinglass}f).
\par Another indicator of the glassy state is the dependence of the decay rate on the amount of time for which the field is applied (Fig. \ref{fig:spinglass}e). This is because, when a field is applied for longer to a spin glass, the spins have more time to settle into a particular ground state, and more spin flips are required to settle into a lower-energy configuration after the field is removed. This effect is most pronounced for $x\sim10\%$ at a temperature of $5$K, where a 1T field was held for 2, 5, or 10 mins. As the duration of field application is increased, a slower relaxation rate is seen, in line with the spin glass model (Fig. \ref{fig:spinglass}e,g). Together, the FC-ZFC irreversibility and the long-timescale non-equilibrium dynamics seen through TRM support the system entering a glassy regime for $x\gtrsim7\%$.

\begin{figure*}[t]
  \centering
  \includegraphics[width=\textwidth]{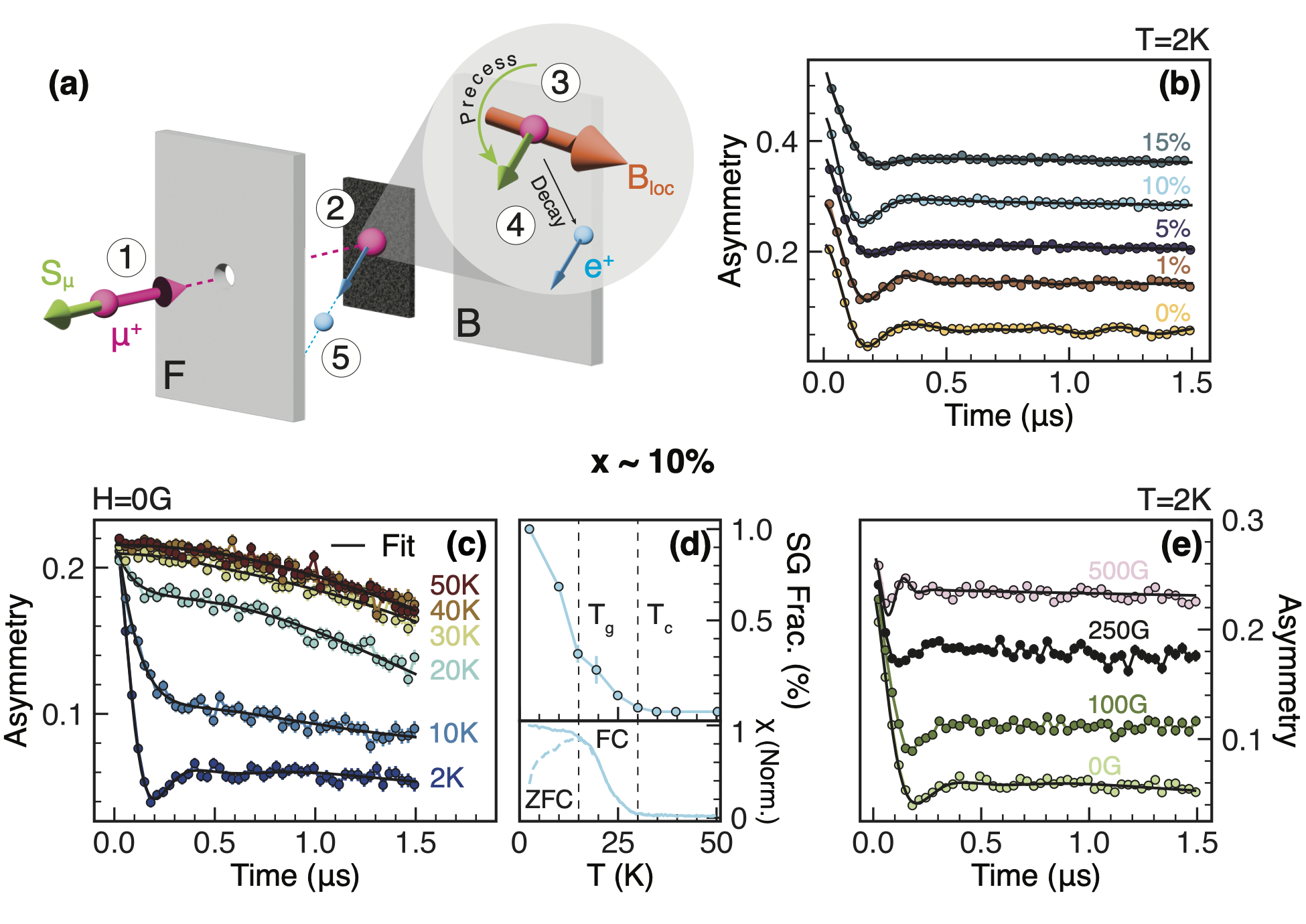}
  \caption{\textbf{Muon Spin Relaxation/Rotation (\musr)} \textbf{(a)} Schematic representation of a typical \musr\ experiment. (1) Spin-polarized muons are sent into the sample chamber, with their spin angular momentum opposite to their direction of motion. (2) A muon implants itself into the material. (3) The spin of the implanted muon precesses in the local magnetic field. (4) The muon decays into a positron, which is emitted preferentially in the direction of the muon's spin. (5) The positron is detected by the front or back detectors. The normalized difference between the number of muons detected over time by these detectors is the measured asymmetry. \textbf{(b)} Asymmetry as a function of time after muon implantation for different dopings, collected at 2~K. Coherent oscillations are seen in the pristine and low dopings, consistent with long-range incommensurate order, while overdamped behavior characteristic of a spin glass state is seen for higher dopings. The spectra for subsequent dopings are offset vertically for clarity. \textbf{(c)} Asymmetry spectra for $x\sim10\%$ in zero field collected at different temperatures. \textbf{(d)} From fits to the curves in \textbf{(c)}, the volume fraction of the spin glass state is extracted, showing an upturn at $T_C$ and $T_g$ roughly consistent with the magnetometry data (lower panel). \textbf{(e)} Asymmetry spectra for $x\sim10\%$ at 5K as a function of applied longitudinal field, showing a steady recovery of asymmetry with increasing field as expected for a static spin glass ground state.}
  \label{fig:muSR}
\end{figure*}
\begin{figure*}[t]
  \centering
  \includegraphics[width=\textwidth]{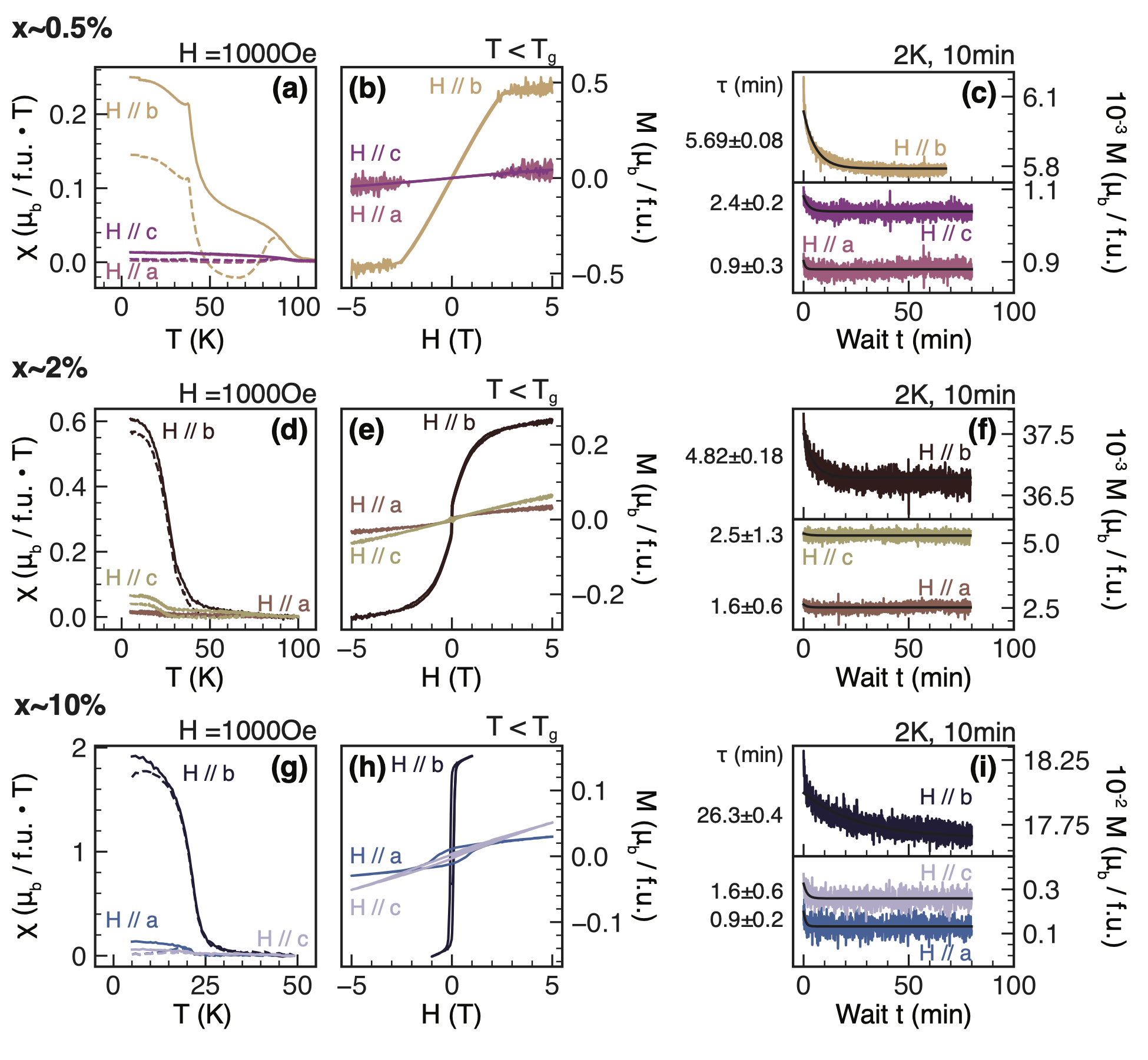}
  \caption{\textbf{Evidence of Spin Glass Anisotropy} \textbf{(a)} Magnetic susceptibility of $x\sim9\%$ with $H=1000$ Oe applied along $a,b,c$, with the $b$-axis remaining the easy axis even in the spin glass state. \textbf{(b)} TRM for the same sample with $H=1$T applied along $a,b,c$ at $T=2K$ for 10min. Exponential fits show that when the field is applied along $b$, the system takes longer to stabilize into a magnetic ground state. \textbf{(c)} Proposed phase diagram for \rulio\ with $x\lesssim10\%$. $T_\eta$ is extracted from the upturn in $\chi_{FC-ZFC}$, while $T_N/T_g$ is extracted by inspecting $d\chi/dT$.}
  \label{fig:anisotropy}
\end{figure*}

\subsection{A Local Probe: Muon Spin Relaxation/Rotation (\texorpdfstring{$\mu$SR}{mSR})}
\par While bulk magnetometry, thermoremanent relaxation, and $ac$-heat capacity measurements establish the emergence of a glassy frozen regime, these probes alone cannot determine whether the underlying magnetic state consists of static random local fields, fluctuating short-range correlations, or residual phase-separated magnetic order. To directly probe the local magnetic environment and distinguish between coherent long-range order and static disordered freezing, we therefore turn to a powerful local probe, muon spin relaxation/rotation (\musr).
\par As large sample volumes were required, polycrystalline powder samples were utilized. Characterization of these samples using powder X-ray diffraction is shown in Fig. 9. In a \musr\ experiment conducted in zero applied field (ZF), shown schematically in Fig. \ref{fig:muSR}a, the time-dependent asymmetry $a(t)$ takes characteristic shapes depending on the nature of the magnetic state~\cite{uemur;ms99}. For long-range-ordered magnetic ground states, the ZF asymmetry shows coherent oscillations with time due to the presence of well-defined local fields at the muon stopping sites, where the width of the field distribution is proportional to the damping rate of the oscillations. This is observed for the ZF asymmetry spectra collected from the $x=0\%$ and $x\sim1\%$ samples at 2K, shown in Fig.~\ref{fig:muSR}b. Both spectra can be fit well using a sum of zeroth-order Bessel functions of the first kind, which is the expected lineshape for incommensurate magnetic order~\cite{blund;cp10}, exactly consistent with the known magnetic ground state in the low doping limit. The faster damping rate in the $x\sim1\%$ sample points to a broader field distribution at the muon stopping sites compared to the undoped sample, attributable to dopant-induced disorder. In contrast, the spectra for $x\gtrsim5\%$ (Fig.~\ref{fig:muSR}b) show only an initial dip in asymmetry followed by a gradual recovery. This overdamped Kubo–Toyabe-like relaxation is characteristic of static random local magnetic fields and is commonly associated with spin-glass freezing \cite{uemur;ms99}, and arises from the distribution of static, random fields at the muon stopping sites. The spectra can be well-fit by the sum of two KT functions. When a longitudinal field (LF) is applied, the initial decrease in asymmetry is suppressed, as seen in Fig.~\ref{fig:muSR}e for the $x\sim10\%$ sample. This ``decoupling'' of the asymmetry demonstrates that the ZF relaxation is due to static magnetic fields as opposed to dynamically fluctuating fields, again consistent with a spin glass ground state.Taken together, the ZF and LF \musr\ spectra demonstrate a continuous doping-driven evolution from coherent incommensurate magnetic order to a frozen spin-glass ground state with static, random local fields.
\par Investigating the temperature-dependent transition into the spin glass ground state is also illuminating. We show this for the $x\sim10\%$ compound in Fig.~\ref{fig:muSR}(c), where representative ZF spectra collected between 2~K and 50~K are displayed. At high temperature, the spectra exhibit slow, Gaussian-like relaxation, consistent with a paramagnetic state. As the temperature decreases, a fast front-end relaxation appears, indicating the presence of static magnetic fields at the stopping positions of some of the muons. The growth of this fast front end is proportional to the fraction of the sample volume with static magnetic fields, i.e. the volume fraction of the spin glass state. At sufficiently low temperatures, the Gaussian-like component is completely eliminated, leaving only the KT-like spin glass relaxation, indicating that nearly the full sample volume participates in the frozen magnetic state. The temperature dependence of the spin glass fraction extracted from our fits to the \musr\ spectra is shown in Fig.~\ref{fig:muSR}(d), demonstrating a gradual, volume-wise transition from the paramagnetic state to the spin glass ground state. The temperature range of the transition is consistent with the magnetometry data. Combined with the absence of competing long-range magnetic order in REXS, these results indicate that the frozen state emerges through disorder-driven destabilization of the incommensurate magnetic manifold rather than reconstruction into an alternative ordered phase.

\section{Discussion}
\par The emergence of spin glass behavior upon doping a Kitaev interaction-dominant system is not unusual. In \runio, doping of $x=5\%$ and above induces a spin glass \cite{mehlawat_fragile_2015}. In A$_{2}$Ti$_{x}$Ir$_{1-x}$O$_{3}$ (A=Na,Li), a spin glass also emerges in both compounds with the lowest levels of nonmagnetic Ti substitution \cite{manni_effect_nonmagnetic_2014}. These studies emphasize the importance of short-range coupling in \nio, as the magnetic order is easily perturbed by magnetic and non-magnetic impurities; even in the case of doping Li on the cation sites of \nio, the impact on the lattice parameters sufficiently disorders the local exchange interactions to drive the system into a spin glass state \cite{rolfs_spiral_2015}. However, in \lio, even though a spin glass does emerge for low levels of Ti, $\Theta_{CW}$ is unaffected, indicating that impurities do not disrupt the exchange interactions to the same extent as in \lio\ \cite{manni_effect_nonmagnetic_2014}. 
\par This means that in \lio, though the magnetically ordered state is extremely fragile to perturbation, the long-range exchange pathways are robust. The spin glass state that emerges from magnetic and non-magnetic impurity doping appears to preserve substantial short-range frustrated correlations while preventing the establishment of coherent long-range magnetic order. In \rulio, this leads to a unique anisotropic quality in the resulting spin glass (Fig. \ref{fig:anisotropy}). For the lowest dopant concentration $x\sim0.5\%$, the expected anisotropy in magnetic susceptibility and saturation magnetization is seen when $H//b$ compared to $H//a,c$ (Fig. \ref{fig:anisotropy}a,b). Furthermore, dynamical timescales are about 2-5$\times$ greater for $H//b$ (Fig. \ref{fig:anisotropy}c). This trend continues through $x\sim2\%$ (Fig. \ref{fig:anisotropy}d-f). In the mid-doping, glassy regime at $x\sim10\%$, this anisotropy becomes \textit{more} pronounced; the low-temperature susceptibility is $\sim30\times$ greater when $H//b$ than $H//c$, as compared to the $\sim20\times$ increase for $x\sim0.75\%$. Furthermore, non-equilibrium dynamical behavior is extremely anisotropic, with a more than $16\times$ increase in the characteristic timescale when $H//b$.
\par This anisotropy can be compared to the case of \rucrcl, where $S=3/2$ moments from Cr$^{3+}$ replace the $J_{eff}=1/2$ moments of the Ru$^{3+}$ ions. For dilute Cr doping, a reversal of the magnetic anisotropy is seen, such that the easy axis moves from the direction perpendicular to the basal plane to the basal plane \cite{bastienSpinglassStateReversed2019a}. The anisotropy seen in \rucrcl\ can be explained by the ferromagnetic Heisenberg-type Cr-Cr exchange interactions, which would prefer to align to an out-of-plane applied field. Additionally, in $\alpha-$Ru$_{1-x}$Ir$_x$Cl$_3$, increased disorder reduces anisotropic response \cite{do2018}. By comparison, the anisotropy present in the spin glass regime of \rulio\ is unique for its enhancement with disorder, and because it reflects the anisotropy of the pristine system, maintaining the $b$-axis as its easy axis. The $b$-axis in \blio\ is a known Kitaev direction, indicating that the bond-directional exchange frustration is maintained in the glassy phase. This is in agreement with the notion that in \lio, long-range exchange pathways are robust to local perturbations. Furthermore, the properties of the glass itself maintain the pristine \lio\ anisotropy, as can be seen in the TRM (Fig. \ref{fig:anisotropy}i). In applying and subsequently removing a field of 1T to the crystalline $a,b,c$-axes, a significantly slower characteristic timescale $\tau$ for the relaxation is seen when the field is applied along the $b$-axis. This indicates that the spins participating in the spin glass are the same as those responsible for the anisotropic behavior.
\par Anisotropy in a spin glass, while not unprecedented \cite{roux-buissonAnisotropicSpinGlass1980,liAnisotropicSpinglassMagnetic2022,yinAnisotropicSpinGlass2024}, is unusual, and to our knowledge this is the first time spin glass anisotropy has been seen in a compound with Kitaev-like interactions. In conventional disorder-driven spin glasses, random exchange interactions tend to average out directional magnetic information. The persistence of strong $b$-axis anisotropy in both the static susceptibility and nonequilibrium relaxation dynamics therefore indicate that the underlying bond-directional exchange network remains partially intact deep within the frozen regime.

\par One interpretation of the above results can be viewed in terms of the $JK\Gamma$ model. In the pristine \blio\ system, the competition between Kitaev, Heisenberg, and off-diagonal exchange interactions stabilizes the incommensurate magnetic ground state. The suppression of $T_N$ with increasing Ru substitution suggests that the interactions responsible for coherent long-range ordering are progressively weakened. The emergence of spin-flop-like behavior near $x\sim5\%$ is likewise consistent with a redistribution of the competing exchange balance within the frustrated manifold. However, rather than stabilizing a coherent quantum-disordered state, the addition of dilute disorder instead freezes the frustrated anisotropic exchange network into a bulk spin-glass phase. By this interpretation, increasing disorder suppresses the interactions responsible for coherent long-range ordering more rapidly than the underlying bond-directional exchange anisotropy, pushing the system toward a more frustrated Kitaev-dominated regime. However, before a coherent quantum-disordered state can emerge, disorder instead stabilizes a frozen anisotropic spin-glass phase around $x\sim7\%$.

\par It is reasonable to assume that all exchange couplings, including $K$, would be weakened by dilute doping; however, previous doping experiments do indicate that the long-range Kitaev correlations are robust to perturbations, and therefore may not be as impacted as the short-range Heisenberg-like interactions $J$. The anisotropy of the Kitaev spin glass further points to remnant Kitaev interactions, with their frustration relieved by the random, dilute doping, creating a glass. These results are consistent with a scenario in which bond-directional exchange anisotropy remains comparatively robust even after the suppression of coherent long-range magnetic order. 
\par It should also be noted that this description does not account for the specific interactions between Ru-Ir and Ru-Ru sites, and at present, the oxidation state of the Ru ions is not known. If the Ru takes a non-magnetic oxidation state, the formation of dilute non-magnetic regions could be sufficient to drive the system into a spin glass. Unfortunately, the insulating nature of the material makes X-ray absorption experiments to verify this valence state challenging, and EDS is not sensitive to low Ru content. However, work by Lei and colleagues supports an Ru$^{4+}$ oxidation state, which is assumed in our interpretation \cite{lei_structural_2014}. The behavior of $S=1$ Ru$^{4+}$ ions on a honeycomb lattice itself is non-trivial, with theoretical modeling \cite{jackeliClassicalDimersDimerized2008} and experimental studies \cite{kimberValenceBondLiquid2014} indicating that the Ru$^{4+}$ sites dimerize via orbital overlap due to orbital degeneracy to form a valence bond liquid, though single-crystalline studies have also shown an antiferromagnetically ordered state \cite{wangLatticetunedMagnetismRu2014}. Whether the Kitaev spin glass in \rulio is driven by non-magnetic or magnetic impurity doping, its effect appears to impact magnetic coupling strengths disproportionally, so that the Kitaev coupling and its resulting anisotropy are maintained. It would be an exciting avenue to explore both of these possibilities experimentally and theoretically. On the experimental end, REXS under applied magnetic field may be conducted on $x\sim5\%$ to better position the strength of relative exchange couplings, as in \cite{ruiz2017correlated}; and and, resonant inelastic X-ray scattering (RIXS) can show if changes in the magnetic excitation spectrum are in agreement with reduced non-Kitaev coupling strengths \cite{ruizMagnonspinonDichotomyKitaev2021a}. Spectroscopy studies may also be attempted to verify the oxidation state of the Ru sites. Theoretical modeling may also reveal the potentiality for new emergent phenomena in a strongly correlated, anisotropic Kitaev spin glass.

\begin{figure*}[ht]
  \centering
  \includegraphics[width=3.6in]{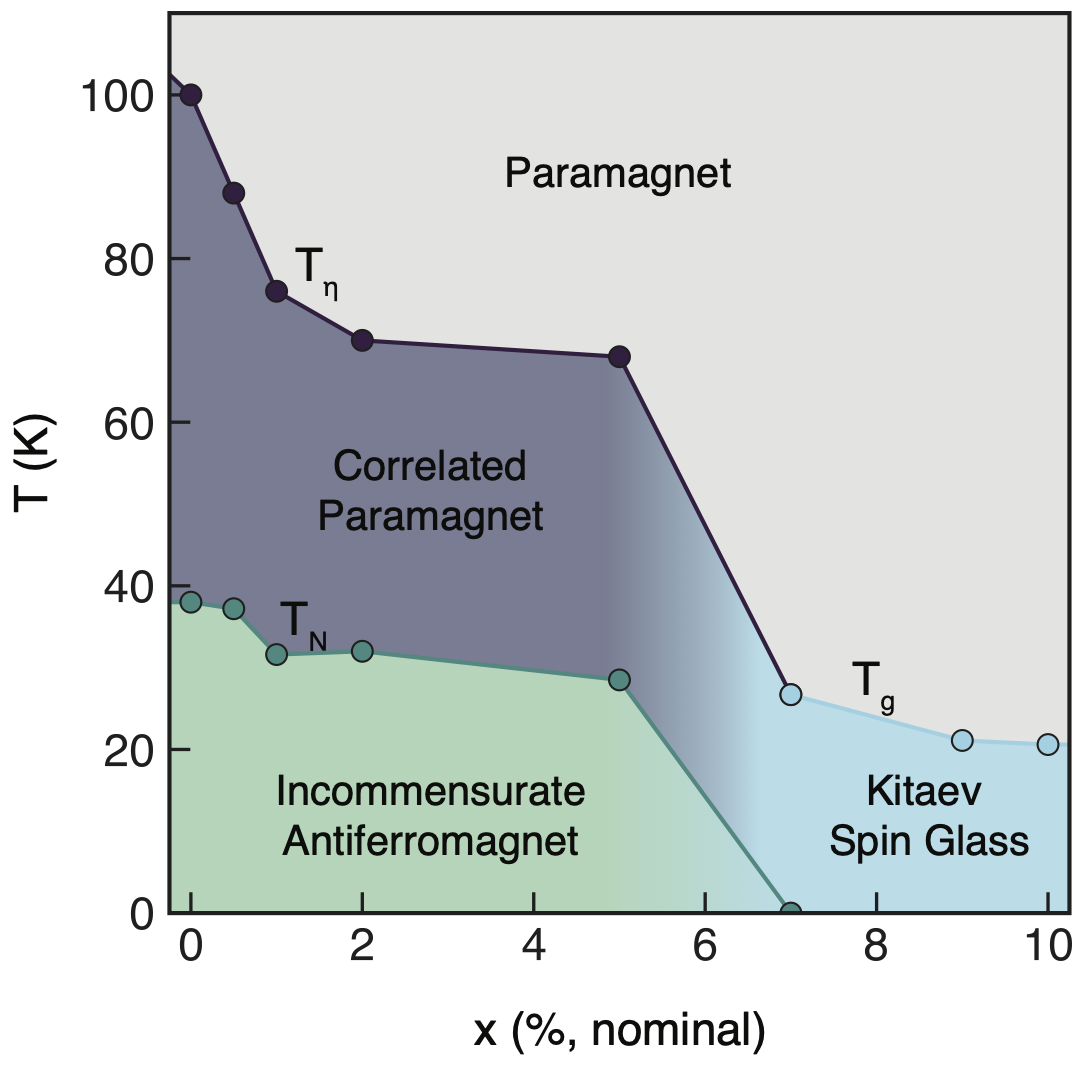}
  \caption{\textbf{Phase Diagram for \rulio for $x\lesssim10\%$} A phase diagram is constructed based on the evolution of $T_N$, $T_{\eta}$, and $T_g$ in \rulio. The low-doping regime hosts the known incommensurate antiferromagnetic ground state of pristine \blio, which is continuously suppressed with increasing disorder. No competing long-range zigzag or stripy magnetic order is observed as the incommensurate phase collapses. Instead, the system evolves into a bulk anisotropic spin-glass state characterized by frozen local magnetic fields, slow relaxation, and strong remnant $b$-axis anisotropy. The higher-temperature crossover scale $T_\eta$ is extracted from the onset of FC–ZFC irreversibility, while $T_N$ and $T_g$ are determined from anomalies in $d\chi/dT$. The resulting phase diagram suggests that dilute disorder destabilizes coherent long-range magnetic order more efficiently than it destroys the underlying bond-directional exchange network.}
  \label{fig:phasediagram}
\end{figure*}

\section{Conclusion}
\par In conclusion, we used a combination of magnetometry, $ac-$heat capacity, resonant elastic X-ray scattering, and \musr\ to investigate the disorder-driven evolution of magnetism in $\beta$-\rulio. We show that dilute Ru substitution continuously suppresses the incommensurate antiferromagnetic ground state of pristine \blio\ without stabilizing an alternative long-range ordered magnetic phase. Instead, the system evolves into a bulk static spin-glass state characterized by slow relaxation, aging behavior, and frozen random local magnetic fields. A phase diagram for \rulio\ for $x\lesssim10\%$ is constructed from these behavior (Fig. \ref{fig:phasediagram}).
\par Remarkably, the frozen regime preserves the strong directional magnetic anisotropy of pristine \blio, including enhanced $b$-axis susceptibility and highly anisotropic thermoremanent relaxation dynamics. Combined reciprocal-space and structural measurements further demonstrate that the hyperhoneycomb lattice remains intact while magnetic coherence collapses, indicating that dilute disorder destabilizes coherent long-range magnetic order more efficiently than it destroys the underlying bond-directional exchange network. Taken together, these results establish \rulio as an anisotropic Kitaev spin glass: a disorder-stabilized frozen state that retains strong signatures of the frustrated exchange geometry underlying the parent Kitaev-proximate magnet.

\section{Acknowledgments}
\par We thank Nicolas Ducharme, Alec Petersen, and Alex Shaw for assistance with the \musr\ data collection. We also thank Yi-Zhuang You for useful discussions. We thank Kohtaro Yamakawa and Andrew J. Gubser for assistance in data retrieval. This material is based upon work supported by the National Science Foundation under Grant No. DMR-2145080. M.A.V. acknowledges support from the National Science Foundation Graduate Research Fellowship under Grant No. DGE-2038238. This research used beamline 4-ID of the National Synchrotron Light Source II, a U.S. Department of Energy (DOE) Office of Science User Facility operated for the DOE Office of Science by Brookhaven National Laboratory under Contract No. DE-SC0012704. Use of the APS was supported by the US Department of Energy, Office of Science, Basic Energy Sciences, under Contract No. DE-AC02-06CH11357. This research used resources of the Advanced Light Source, a U.S. DOE Office of Science User Facility under contract no. DE-AC02-05CH11231.

\bibliography{references}
\clearpage

\setcounter{figure}{0}
\renewcommand{\thefigure}{SI\arabic{figure}}

\setcounter{table}{0}
\renewcommand{\thetable}{SI\arabic{table}}

\setcounter{equation}{0}
\renewcommand{\theequation}{SI\arabic{equation}}
\setcounter{section}{0}
\renewcommand{\thesection}{SI\arabic{section}}

\begin{center}
{\Large\bfseries Supplementary Information}
\end{center}
\par The Supplementary Information provides additional experimental details supporting the disorder evolution discussed in the main text. Sections I and II establish the relative Ru concentration hierarchy and sample characterization. Figs. 3–7 track the evolution of resonant magnetic scattering across the doping series, demonstrating the progressive suppression of coherent incommensurate order. Fig. 8 summarizes the structural evolution and shows that the hyperhoneycomb lattice remains intact throughout the explored doping range. Fig. 9 shows powder X-ray diffraction refinement for several samples, confirming the presence of multiple phases in \lio\ powders.

\section{Determination of Relative {Ru} Concentration}
\begin{figure*}[ht]
\includegraphics[width=0.8\textwidth]{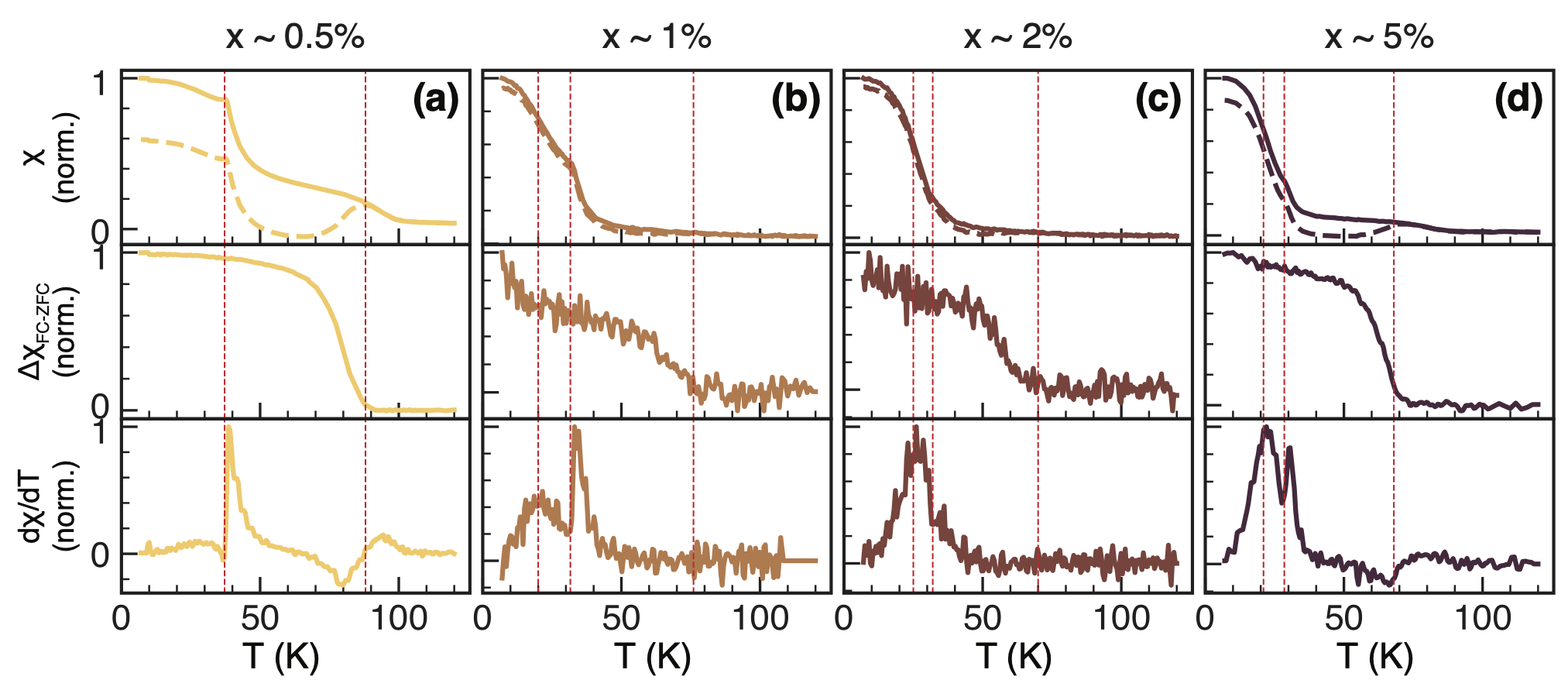}
  \caption{Normalized magnetic susceptibility $\chi$, FC--ZFC irreversibility $\Delta\chi=\chi_{FC}-\chi_{ZFC}$, and $d\chi/dT$ for representative low-doping samples used to establish the relative Ru concentration hierarchy. Systematic suppression of the magnetic energy scales together with the evolution of low-temperature irreversibility provides a phenomenological framework for assigning relative Ru concentrations in the dilute-doping regime where direct compositional determination is challenging.}
  \label{fig:lowdoping}

\includegraphics[width=0.8\textwidth]{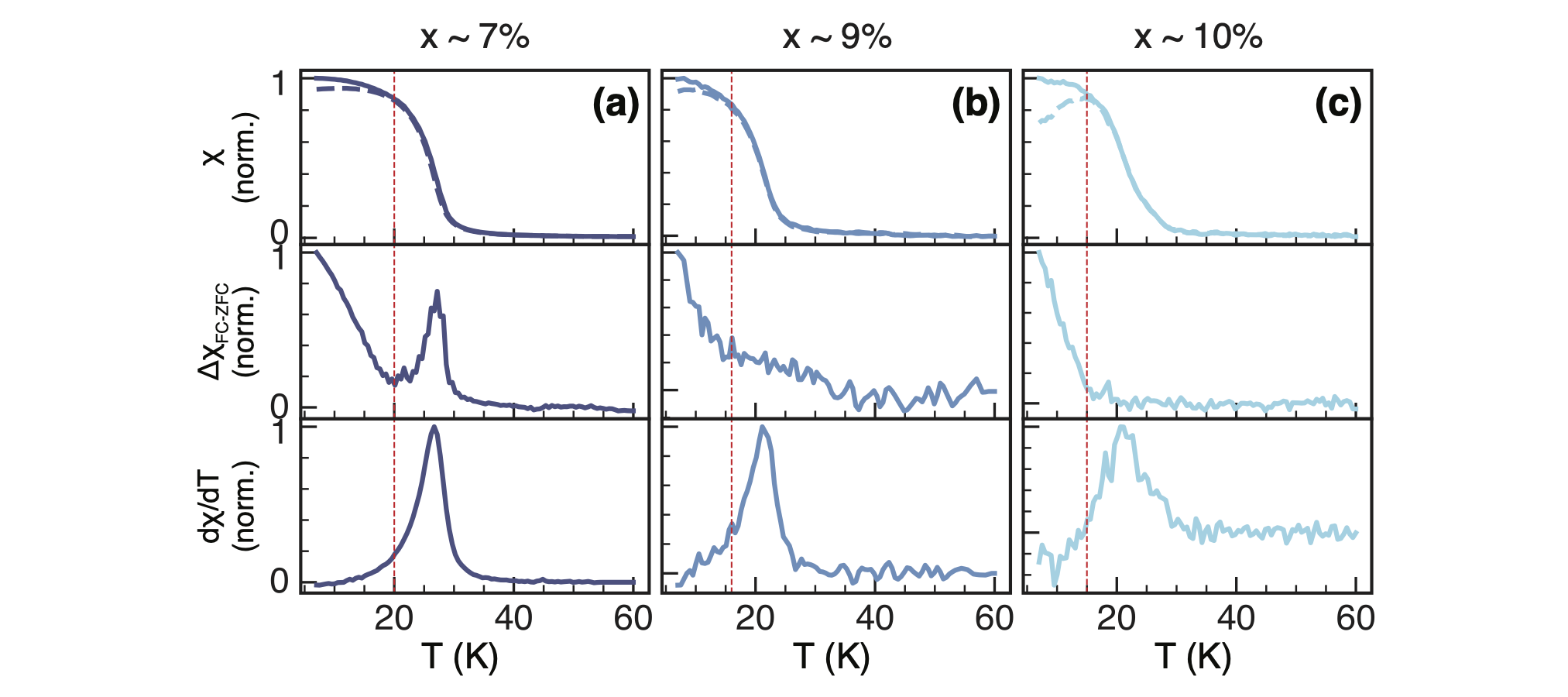}
  \caption{Normalized magnetic susceptibility $\chi$, FC--ZFC irreversibility $\Delta\chi=\chi_{FC}-\chi_{ZFC}$, and $d\chi/dT$ for representative higher-doping samples. The disappearance of sharp order-like features together with the emergence of broad FC--ZFC splitting and low-temperature irreversibility indicates the evolution from coherent incommensurate magnetic order toward glassy freezing with increasing Ru substitution. Dashed red lines mark the operationally defined spin-glass freezing temperature $T_g$.}
  \label{fig:middoping}
\end{figure*}

\par One of the chief challenges in a dilute doping study is the determination of dopant concentration. At low Ru concentrations ($x < 10\%$), energy-dispersive X-ray spectroscopy (EDS) cannot reliably resolve the dopant level within the small single crystals studied here. Additionally, the insulating behavior of the honeycomb iridates makes direct determination of the Ru concentration using total-electron-yield X-ray absorption spectroscopy particularly challenging. Because of these limitations, we discuss throughout the main text not an absolute Ru concentration, but a relative concentration hierarchy established phenomenologically from systematic magnetic evolution. Previous studies on polycrystalline samples provide an approximate reference scale for the accessible doping regime \cite{lei_structural_2014}.

To establish this hierarchy, we compare the systematic evolution of several experimentally measured quantities, including the suppression of $T_N$ and $T_\eta$, the appearance and growth of $T_g$, the low-temperature susceptibility, and the saturation magnetization. Representative low- and high-doping susceptibility systematics are summarized in Figs. \ref{fig:lowdoping}-\ref{fig:middoping}. At low doping, the first derivative of $\chi_{\mathrm{ZFC}}$ exhibits a well-defined feature associated with $T_N$, while increasing Ru substitution progressively suppresses this anomaly and enhances low-temperature irreversibility. In the higher-doping regime, the sharp order-like feature disappears and only a broad low-temperature freezing anomaly remains. Together, these trends provide a consistent phenomenological framework for assigning relative Ru concentrations across the dilute-doping series.

\def\arraystretch{1.25}
\begin{table}
\begin{center}
\begin{tabular}{|c|c|c|c|} 
 \hline
 $x$ (nominal) & $T_N$ & $T_{\eta}$ & $T_g$\\
 \hline
 0.5\% & 37.2K & 88K & -\\
 \hline
 1\% & 31.6K &76K  & -\\
 \hline
 2\% & 32.0K & 70K & -\\
 \hline
 5\% & 28.5K & 68K & - \\
 \hline
 7\% & - & - & 20K \\
 \hline
 9\% & - & - & 16K \\
 \hline
 10\% & - & - & 15K\\
 \hline
 
 \hline
\end{tabular}\caption{$T_N,\,T_{\eta},$ and $T_g$ determined from bulk magnetometry}
\end{center}
\end{table}
Representative structural and resonant magnetic scattering measurements across the dilute Ru-substitution series are shown in Figs. \ref{fig:REXSp5} - \ref{fig:REXS10}. Sharp structural Bragg peaks and narrow rocking curves demonstrate that the hyperhoneycomb lattice remains structurally coherent throughout the explored doping range. At low Ru concentrations, the incommensurate magnetic reflection remains resonantly enhanced and well-resolved, indicating that coherent long-range magnetic order initially survives weak disorder. With increasing Ru substitution, the magnetic reflection progressively weakens and broadens before disappearing entirely at the highest measured concentrations, while no competing zigzag or stripy magnetic reflections are observed within the explored reciprocal-space region. In contrast, the structural reflections remain sharp across the series, indicating that the suppression of magnetic coherence occurs without a corresponding crystallographic phase transition.

The structural evolution of the dilute-doping series is summarized in Tables \ref{tab:REXS_latt}-\ref{tab:microdiff_latt}. The lattice parameters and anisotropy ratios evolve only modestly with increasing Ru substitution, further supporting the conclusion that the dramatic magnetic reconstruction discussed in the main text occurs within an otherwise intact hyperhoneycomb crystal framework.

\par In comparing the systematics of these samples, there are several factors to consider: The movement of $T_N$ and $T_{\eta}$, the magnitude of the low-temperature susceptibility $\chi_{base}$, the saturation magnetization $M_{sat}$, and the appearance and movement of $T_{g}$. A first derivative of $\chi_{ZFC}$ was used to determine $T_N$; the corresponding feature is marked in Figs. \ref{fig:lowdoping},\ref{fig:middoping}. As in \cite{ruiz;prb20}, $T_\eta$ was determined using the upturn in $\chi_{FC}-\chi_{ZFC}$. An additional feature was seen in the first derivative of $\chi_{ZFC}$ at $T_{inc}<T_N$, corresponding to an increase in susceptibility at low temperatures. This may occur to a decrease in strength of AFM interactions, which allows spins to more easily align with the applied field. This peak was also used as a marker for relative dopant concentration. For $x\sim2\%$, the peak in d$\chi$/dT was less pronounced due to the small magnitude of suscpetibiliy; a small upturn in $\Delta\chi$ seen for $x\sim1,5\%$ was instead used to determine $T_N$. Meanwhile, for the three sample with higher doping, only a single feature was seen in d$\chi$/dT, near an exponential increase in $\Delta\chi$. $T_g$ was instead determined by the FC-ZFC splitting which occurs after the upturn in $\chi$, as marked by the dashed red line in Fig.\ref{fig:middoping}.

\section{Sample Details}
\def\arraystretch{1.25}
\begin{table}[h]
\begin{center}
\begin{tabular}{|c|c|c|c|} 
 \hline
 $x$ (\%, nominal) & a (\AA) & b (\AA) & c (\AA) \\
 \hline
 0\% & 5.91 & 8.43 & 17.80 \\
 \hline
 0.5\% & 5.9528 & 8.40593 & 17.7766 \\
 \hline
 1\% & 5.89034 & 8.62265 & 17.7893 \\
 \hline
 2\% & 5.87799 & 8.4261 & 17.7928 \\
 \hline
 10\% & 5.8998 & 8.32549 & 17.7135 \\
 \hline
\end{tabular}\caption{Lattice parameters determined by REXS at APS beamline 6-ID-B and NSLS beamline 4-ID.}\label{tab:REXS_latt}
\end{center}
\end{table}

\par
\begin{figure*}[ht]
\includegraphics[width=0.7\textwidth]{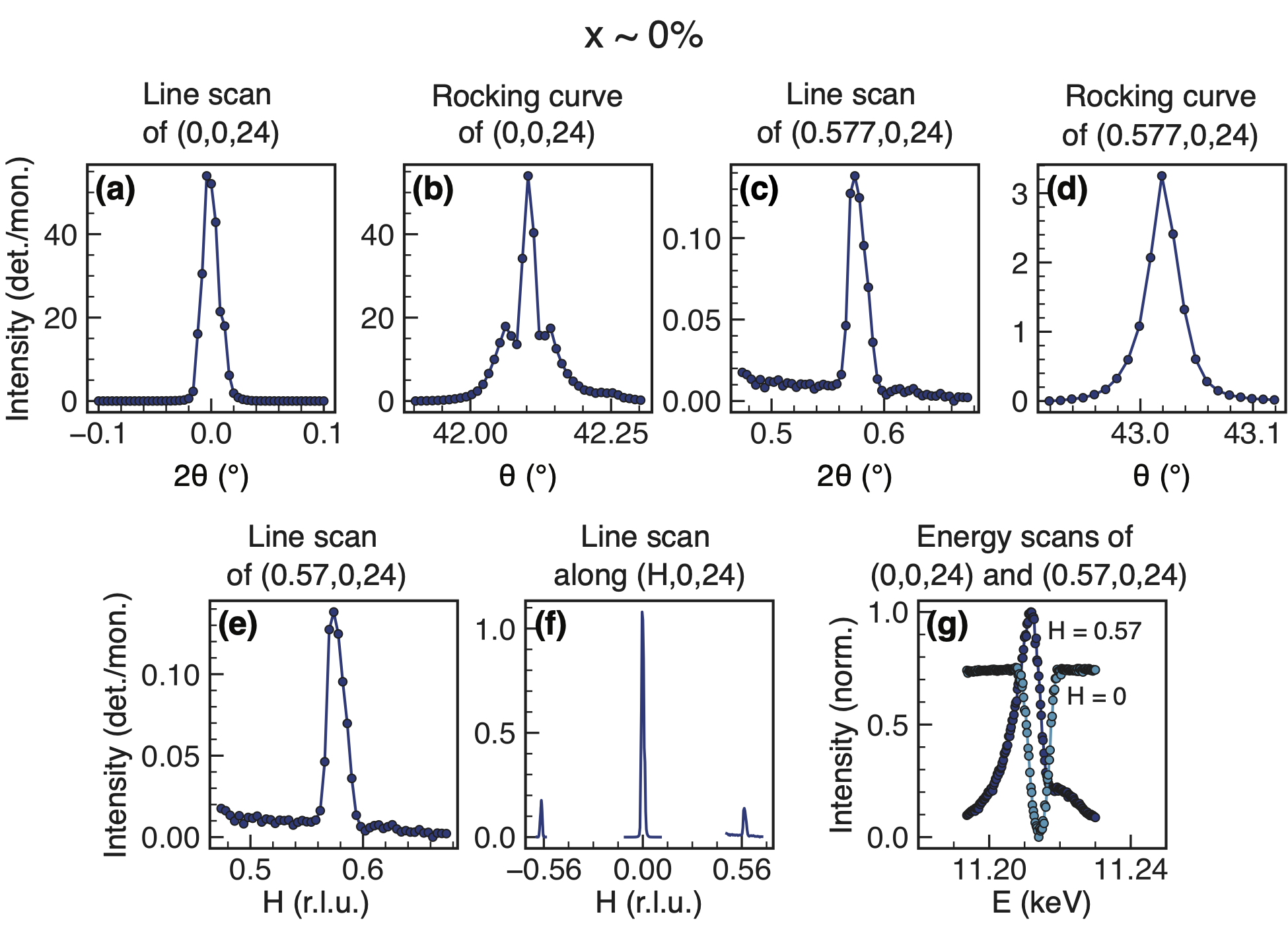}
  \caption{Structural and resonant magnetic scattering measurements for pristine $\beta$-Li$_2$IrO$_3$ ($x\sim0\%$) taken at NSLS II beamline 4-ID. Sharp structural Bragg peaks and narrow rocking curves demonstrate high crystal quality, while the resonantly enhanced incommensurate magnetic reflection confirms coherent long-range magnetic order in the parent compound.}
  \label{fig:REXS0}
\includegraphics[width=0.7\textwidth]{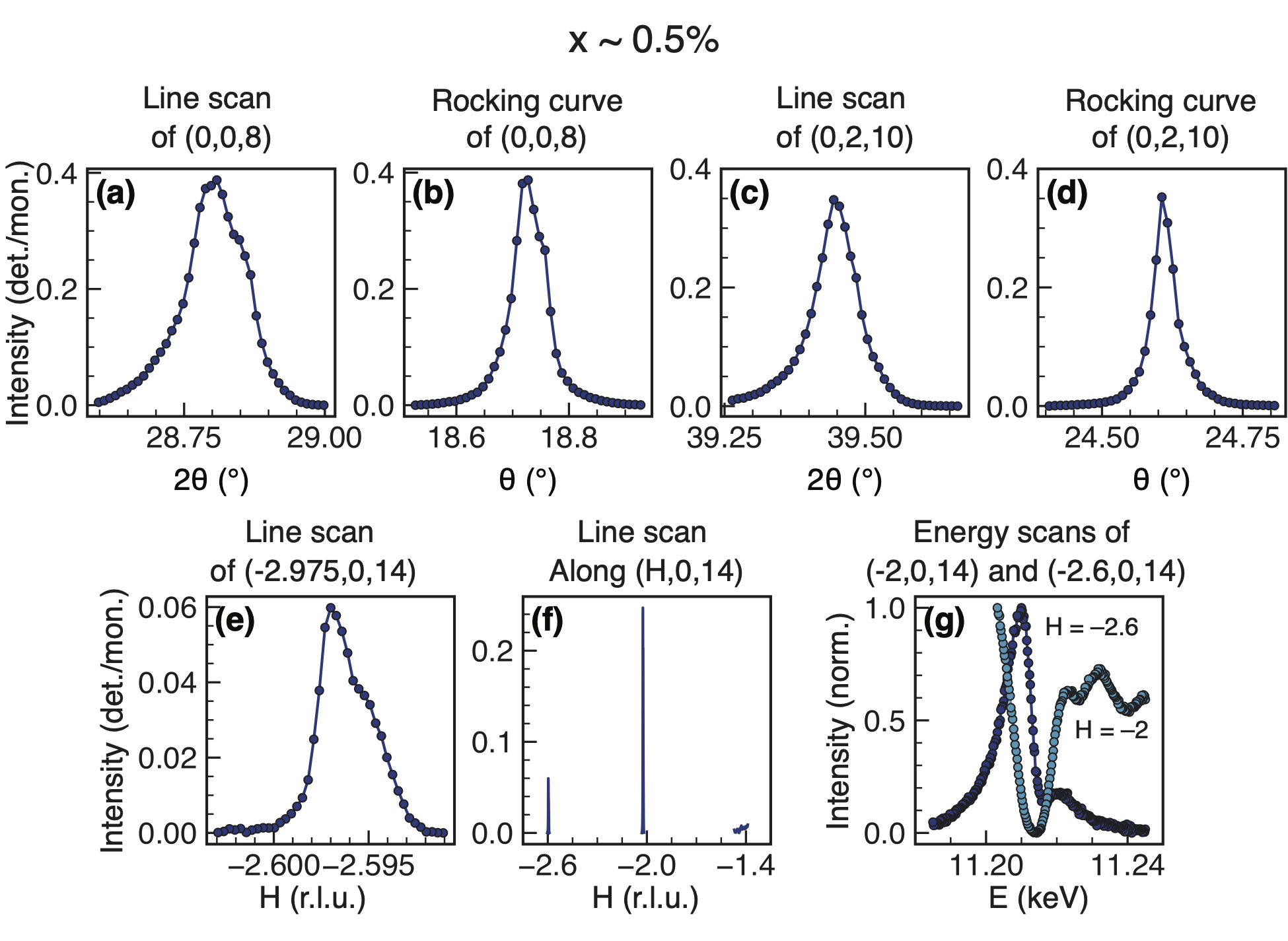}
  \caption{Structural and resonant magnetic scattering measurements for $\beta$-Li$_2$Ru$_x$Ir$_{1-x}$O$_3$ with $x\sim0.5\%$ taken at the APS beamline 6-ID-C. The incommensurate magnetic reflection remains sharp and resonantly enhanced, demonstrating that coherent long-range magnetic order survives at very low Ru substitution while the hyperhoneycomb crystal structure remains intact. An incommensurate peak was not clearly seen at $\vec{q}=(-1.4,0,14)$ due to sample misalignment.}
  \label{fig:REXSp5}
\end{figure*}

\begin{figure*}[ht]
\includegraphics[width=0.7\textwidth]{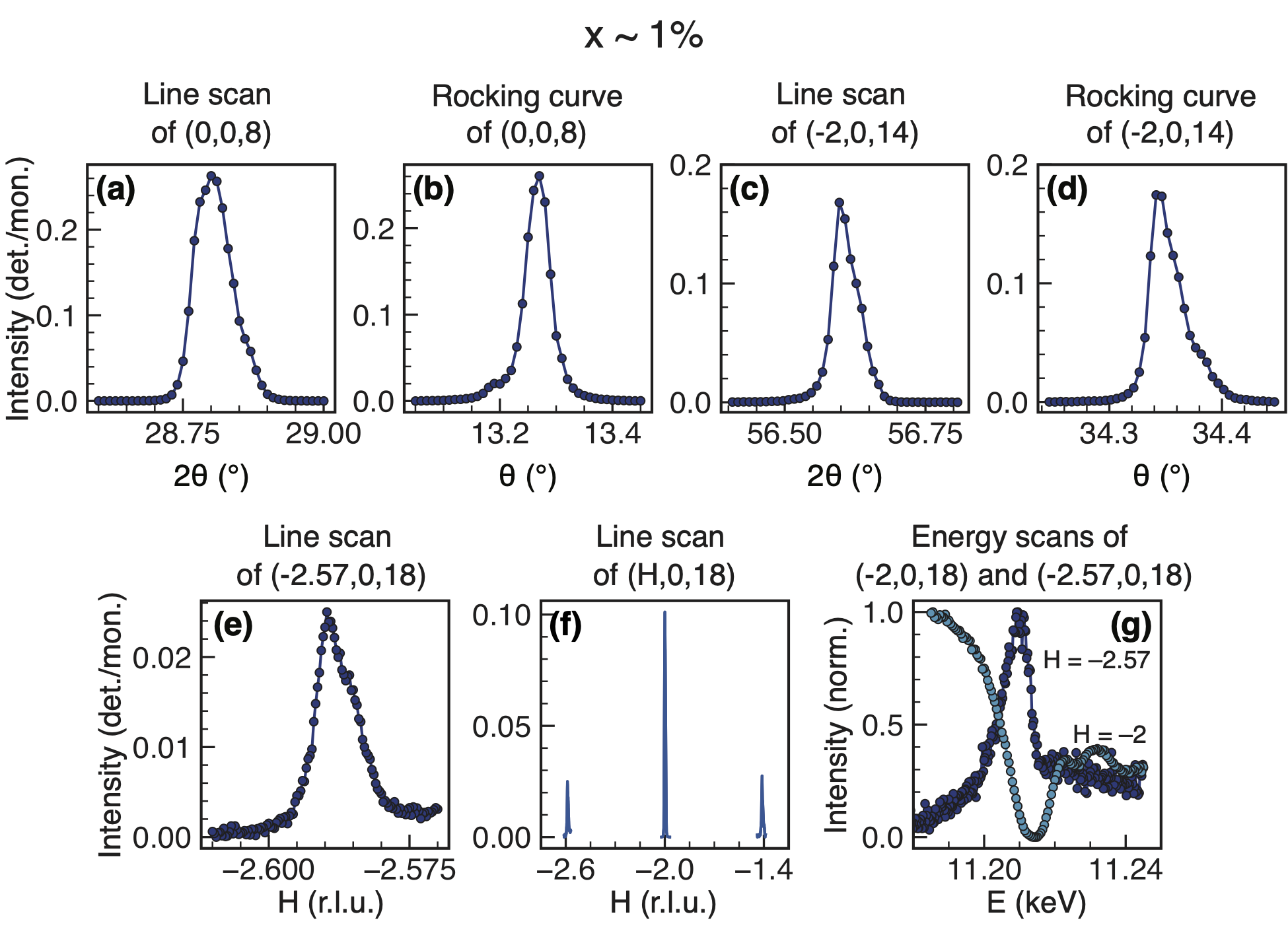}
  \caption{Structural and resonant magnetic scattering measurements for $\beta$-Li$_2$Ru$_x$Ir$_{1-x}$O$_3$ with $x\sim1\%$. The incommensurate magnetic reflection remains observable and resonantly enhanced, but exhibits noticeable weakening and broadening relative to lower doping, indicating progressive suppression of coherent long-range magnetic order while the underlying crystal structure remains intact.}
  \label{fig:REXS1}
  \includegraphics[width=0.7\textwidth]{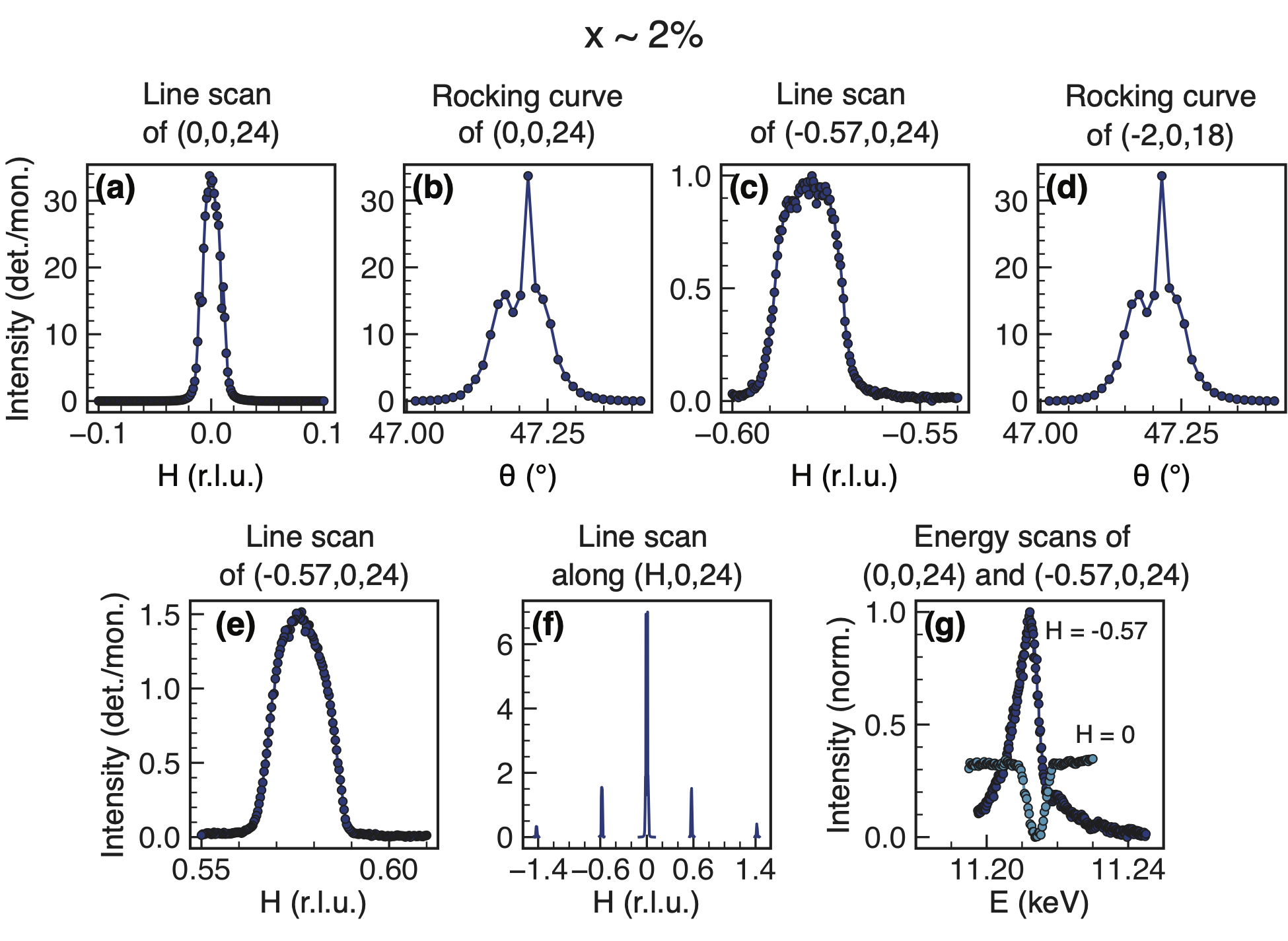}
  \caption{Structural and resonant magnetic scattering measurements for $\beta$-Li$_2$Ru$_x$Ir$_{1-x}$O$_3$ with $x\sim2\%$. The incommensurate magnetic reflection remains detectable and resonantly enhanced, but exhibits substantial weakening and loss of coherence relative to lower doping, indicating that the system approaches the instability of long-range magnetic order while remaining within the same incommensurate magnetic manifold.}
  \label{fig:REXS2}
\end{figure*}

\begin{figure*}[ht]
\includegraphics[width=0.7\textwidth]{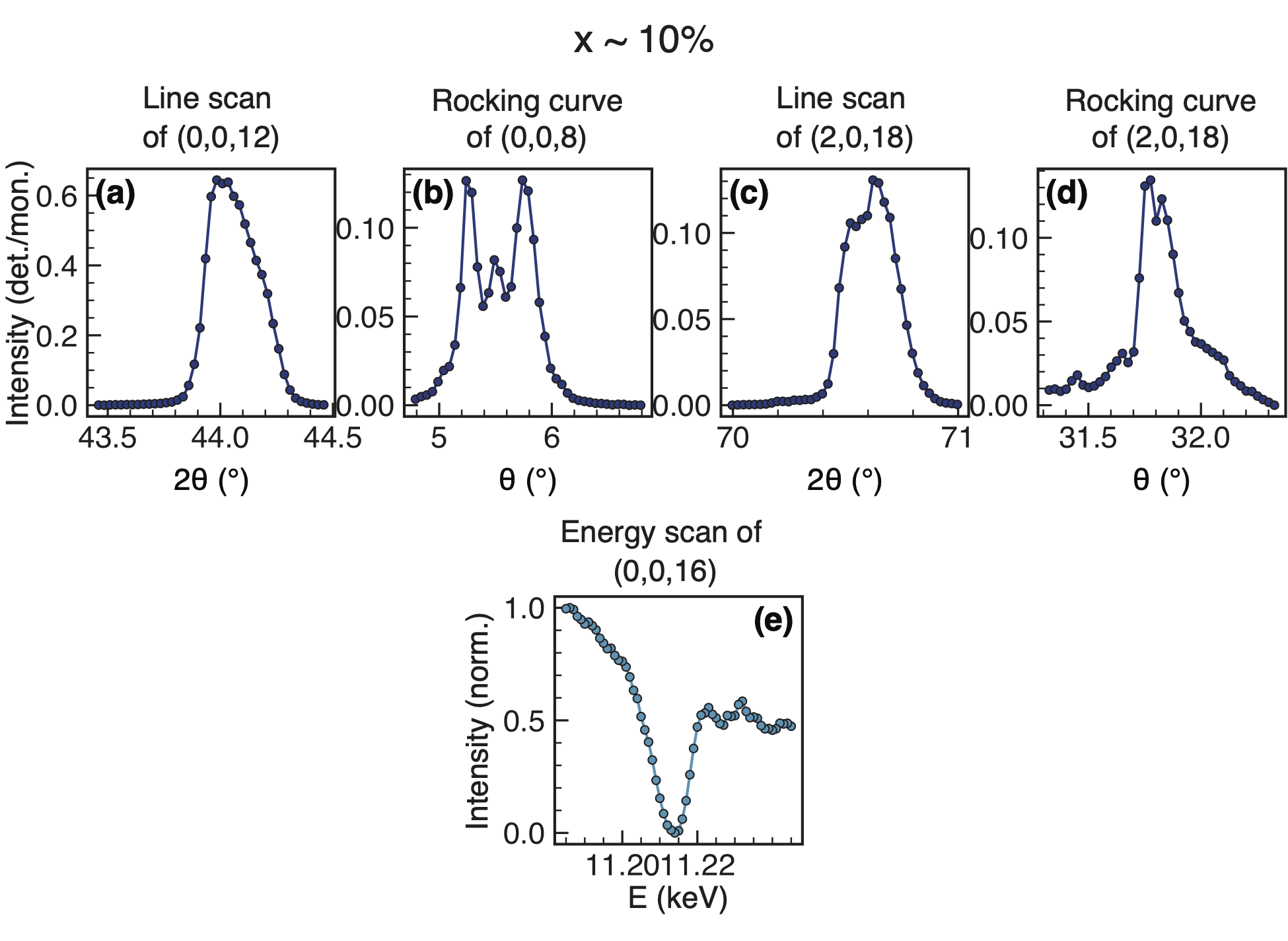}
\caption{Structural and resonant magnetic scattering measurements for $\beta$-Li$_2$Ru$_x$Ir$_{1-x}$O$_3$ with $x\sim10\%$. While sharp structural Bragg peaks remain observable, no coherent incommensurate magnetic reflection is detected, indicating that long-range magnetic coherence is lost even though the hyperhoneycomb crystal structure remains intact.}
  \label{fig:REXS10}
\end{figure*}

 Structural information on \rulio\ single crsytals have been acquired using Laue microdiffraction and REXS. REXS results shown in Figs. \ref{fig:REXS0}-\ref{fig:REXS10} show line scans and rocking curves of in-plane and out-of-plane structural reflections (a-d in each figure), as well as magnetic reflections and their positioning with respect to the nearest structural Bragg peak (Figs. \ref{fig:REXS0}-\ref{fig:REXS2}e-f). Strong absorption is seen in the structural Bragg peaks near the Ir $L_3$ edge, confirming its origin is from the sample, while strong resonant enhancement is seen in the magnetic peaks, indicating magnetism arising from the Ir sites (Fig.\ref{fig:REXS0}-\ref{fig:REXS2}g). SI Table \ref{tab:REXS_latt} shows lattice parameters derived for each sample from REXS experiments. A trend is visible in which $c-$axis lattice parameters increase with increased doping, then decrease in the glassy regime. Furthermore, the positioning of structural and magnetic Bragg peaks confirms the $\beta$-phase is attained for each doping.
\def\arraystretch{1.25}
\begin{table}
\begin{center}
\begin{tabular}{|c|c|c|c|} 
 \hline
 $x$ (nom.) & a (\AA) & (\AA) b (\AA) & c (\AA)\\
 \hline
 0.5\% & 5.9011$\pm$0.0002 & 8.4442$\pm$0.0006 & 17.7958$\pm$0.0012 \\
 \hline
 1\% & 5.9058$\pm$0.0007 & 8.4376$\pm$0.0021 & 17.7964$\pm$0.0028 \\
 \hline
 2\% & 5.9075$\pm$0.0010 & 8.4363$\pm$0.0016 & 17.7932$\pm$0.0011 \\ 
 \hline
 10\% & 5.9169$\pm$0.0007 & 8.4337$\pm$0.0009 & 17.7706$\pm$0.0014 \\
 \hline
 15\% & 5.9307$\pm$0.0015 & 8.4344$\pm$0.0015 & 17.7273$\pm$0.0047 \\
 \hline
\end{tabular}\caption{Lattice parameters determined by Laue microdiffraction at the ALS beamline 12.3.2.}\label{tab:microdiff_latt}
\end{center}
\end{table}

\par
\begin{figure*}[ht]
\includegraphics[width=\textwidth]{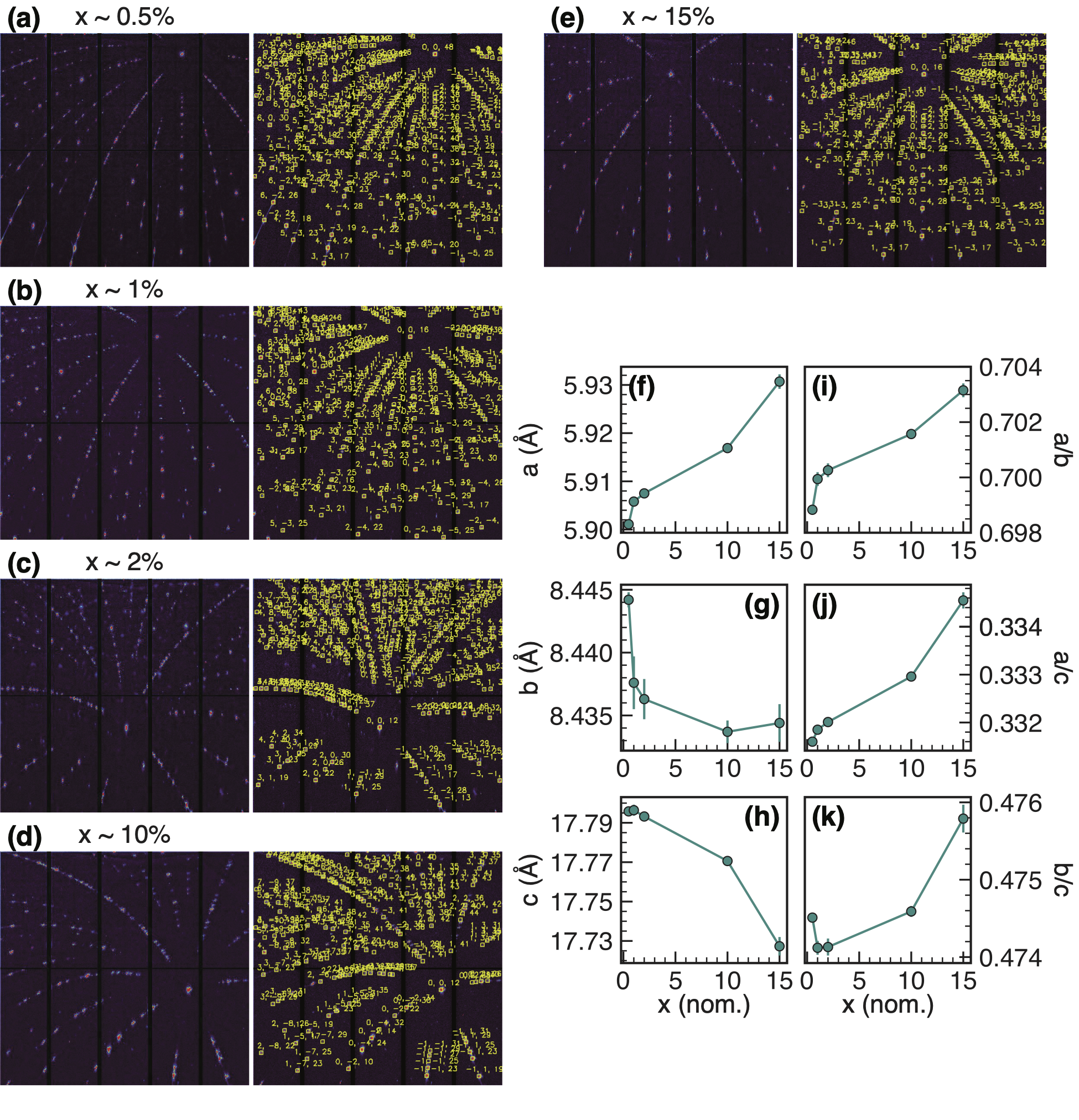}
  \caption{Representative microdiffraction patterns and lattice parameters for $\beta$-Li$_2$Ru$_x$Ir$_{1-x}$O$_3$ across the dilute Ru-substitution series. The hyperhoneycomb crystal structure remains intact throughout the explored doping range, while the lattice parameters and anisotropy ratios evolve only modestly with increasing Ru concentration, indicating that the dramatic magnetic reconstruction occurs without a corresponding structural phase transition.}

  \label{fig:microdiffraction}
\end{figure*}

 Further structural characterization was performed by Laue microdiffraction at ALS beamline 12.0.2 to investigate variation in crystal structure within each sample (Fig. \ref{fig:microdiffraction}). Characteristic microdiffraction scans with background subtraction are shown for each sample, with peak indexing corresponding to an assumed $Fddd$ orthorhombic lattice (Fig. \ref{fig:microdiffraction}d-e). From mapping of the sample, relative lattice parameters were obtained, and absolute parameters were extracted assuming uniform sample volume. Error bars were derived from variation within samples. For the most part, with increasing $x$, a systematic increase is seen in the ratios $a/b$, $a/c$, and $b/c$, indicating that $a$ is increasing with respect to $b,c$ and $b$ is increasing with respect to $c$ for fixed $x$. Assuming fixed unit cell volume for each sample, this corresponds to an expansion in $a$ and a contraction in $b,c$, in agreement with trends seen in REXS.
\par
\begin{figure*}[ht]
\includegraphics[width=\textwidth]{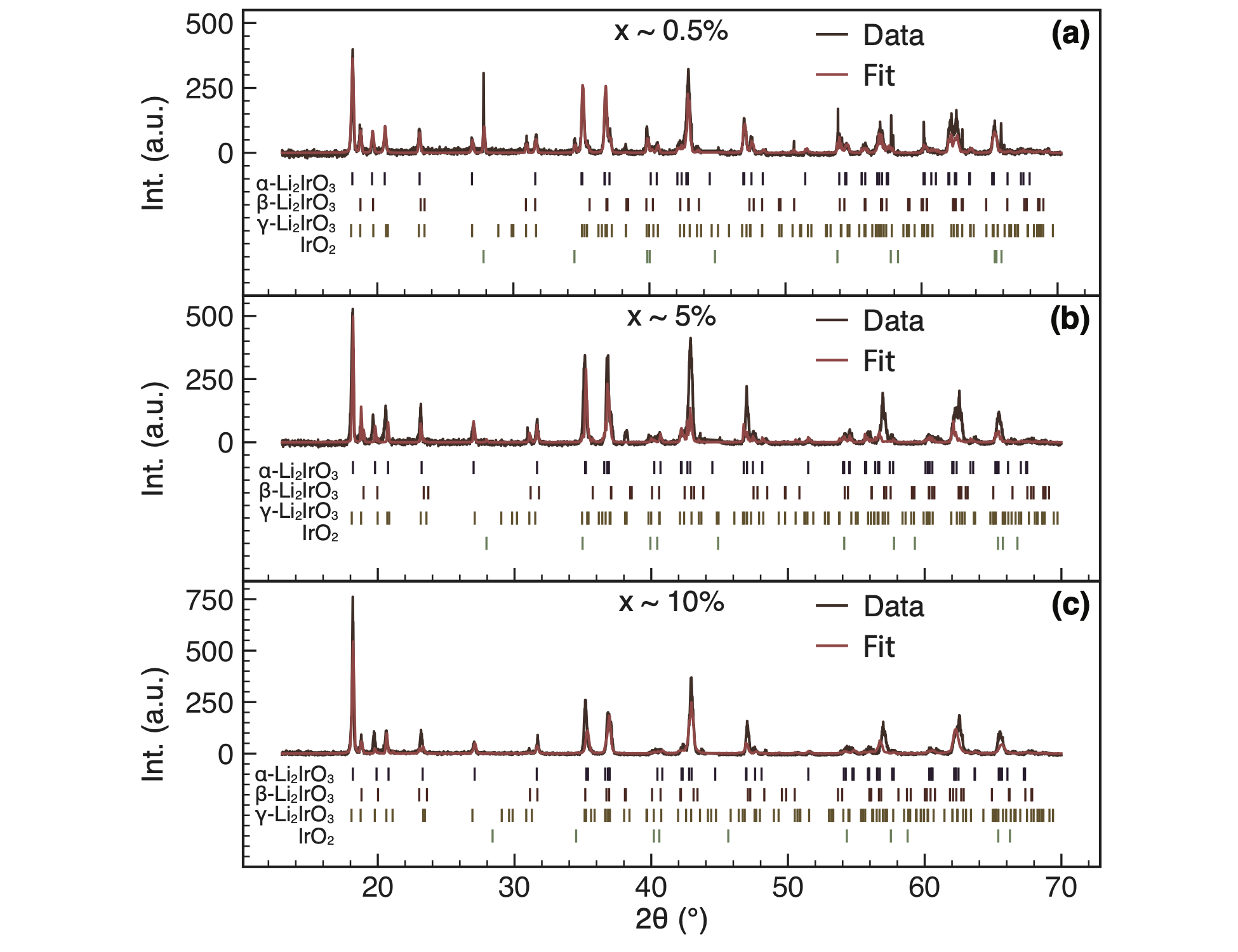}
  \caption{(a-c) Powder X-ray diffraction for powder samples with nominal concentrations $x\sim0.5,5,10\%$, respectively. Red lines correspond to a coarse fit using $\alpha,\beta,$ and $\gamma$ phase of \lio, as well as IrO$_2$. Lines at the bottom of each panel indicate peaks corresponding to each phase. Multiple phases of \lio\ are seen in each sample.}

  \label{fig:pxrd}
\end{figure*}

 Powder X-ray diffraction (PXRD) data were taken for samples used in \musr\ experiments. Each powder is confirmed in preliminary refinement to contain all three phases of \lio, as well as a non-negligible concentration of IrO$_2$. Despite the presence of multiple phases, \musr\ results correspond well with single crystal data. Representative PXRD verifying multi-phase powder is shown in Fig. \ref{fig:pxrd}.

\end{document}